\documentclass[preprint
 amsmath,amssymb,
 aps, physrev,
]{revtex4-2}

\usepackage{graphicx}% Include figure files
\usepackage{dcolumn}% Align table columns on decimal point
\usepackage{bm}% bold math
\usepackage[T1]{fontenc}
\usepackage{hyperref}
\usepackage{amsmath,amssymb,amsfonts}%
\usepackage{longtable}
\usepackage{xcolor}
\begin{document}

\preprint{APS/123-QED}

\title{\textbf{Filtering amplitude dependence of correlation dynamics in complex systems: application to the cryptocurrency market} 
}% 

\author{Marcin W\k{a}torek}
 \email{Contact author: marcin.watorek@pk.edu.pl}
\affiliation{Faculty of Computer Science and Mathematics, Cracow University of Technology, Krak\'ow, Poland}
\affiliation{Adapt Research Centre, School of Computing, Dublin City University, Dublin, Ireland}
\author{Marija Bezbradica}%

\affiliation{Adapt Research Centre, School of Computing, Dublin City University, Dublin, Ireland}

\author{Martin Crane}
 \affiliation{Adapt Research Centre, School of Computing, Dublin City University, Dublin, Ireland}

\author{Jaros{\l}aw Kwapie\'n}
\affiliation{Complex Systems Theory Department, Institute of Nuclear Physics, Polish Academy of Sciences, Krak\'ow, Poland}%

\author{Stanis{\l}aw Dro\.zd\.z}
 \email{Contact author: stanislaw.drozdz@ifj.edu.pl}
\affiliation{Complex Systems Theory Department, Institute of Nuclear Physics, Polish Academy of Sciences, Krak\'ow, Poland}%
\affiliation{Faculty of Computer Science and Mathematics, Cracow University of Technology, Krak\'ow, Poland}

\date{\today}% It is always \today, today,
             %  but any date may be explicitly specified

\begin{abstract}Based on the cryptocurrency market dynamics, this study presents a general methodology for analyzing evolving correlation structures in complex systems using the $q$-dependent detrended cross-correlation coefficient $\rho(q,s)$. By extending traditional metrics, this approach captures correlations at varying fluctuation amplitudes and time scales. The method employs $q$-dependent minimum spanning trees ($q$MSTs) to visualize evolving network structures. Using minute-by-minute exchange rate data for 140 cryptocurrencies on Binance (Jan 2021–Oct 2024), a rolling window analysis reveals significant shifts in $q$MSTs, notably around April 2022 during the Terra/Luna crash. Initially centralized around Bitcoin (BTC), the network later decentralized, with Ethereum (ETH) and others gaining prominence. Spectral analysis confirms BTC’s declining dominance and increased diversification among assets. A key finding is that medium-scale fluctuations exhibit stronger correlations than large-scale ones, with $q$MSTs based on the latter being more decentralized. Properly exploiting such facts may offer the possibility of a more flexible optimal portfolio construction. Distance metrics highlight that major disruptions amplify correlation differences, leading to fully decentralized structures during crashes. These results demonstrate $q$MSTs’ effectiveness in uncovering fluctuation-dependent correlations, with potential applications beyond finance, including biology, social and other complex systems.
\end{abstract}

\maketitle

%\tableofcontents

\section{Introduction}
\label{introduction}
The fundamental characteristic of complex systems is the nonlinear interactions between their constituent elements~\citep{AndersonPW-1972a,MitchellM-2009a}. In such systems, the evolution is typically driven by the presence of multiple generators. As a result, the signals recorded from such systems typically comprise a convolution of effects induced by different generators, where different ones may dominate at different times. This complexity emerges frequently in the form of multifractality, particularly when some of the generators exhibit a hierarchical structure such as in a multiplicative cascade~\citep{BarralJ-2015a}. In such cases, complexity may be encoded in the nonlinear temporal dependencies within the sequence of signal fluctuations, but also in their amplitude, which can strongly be influenced by the temporal dependencies. It is often more pronounced in fluctuations within a certain amplitude range rather than in those outside of it. Typically, fluctuations of relatively large size exhibit a more complex structure compared to small-size fluctuations, which are often overwhelmed by background noise~\citep{DrozdzS-2015a}. From the perspective of identifying complexity in the recordings of a given observable, it is crucial to employ a tool capable of distinguishing data that are relevant to the structural complexity from those that are not. In this regard, signal filtering procedures can be utilized to extract the signatures of complexity, facilitating a more accurate characterization of the underlying dynamical processes. In this work, we apply one such tool based on the multiscale detrended cross-correlation coefficient and graph theory to a structural study of the cryptocurrency market~\citep{AsteT-2019a,ScharF-2020a}.

More than a decade after the introduction of the first cryptocurrency using blockchain technology, Bitcoin, the market for these assets exhibits many characteristics of a mature market. Such characteristics include liquidity~\citep{Corbet2019,WatorekM-2021b,CorbetS-2022a}, power-law tails in probability density distributions~\citep{DrozdzS-2018a,BegusicS-2018a,WatorekM-2021a} and multiscaling~\citep{TakaishiT-2018a,StavroyiannisS-2019a,WatorekM-2021b,KwapienJ-2022a,KwapienJ-2022b,BroutyX-2024a}. In addition to these, similarities with other financial markets has been noted e.g. a level of efficiency~\citep{SensoyA-2019a,TakaishiT-2020a,KakinakaS-2022a} and also significant correlation with such markets~\citep{DrozdzS-2018a,ConlonT-2020a,James2021,James2021b,ZhangYJ-2021a,ElmelkiA-2022a,WatorekM-2023a,Li2025}. This last characteristic prevents cryptocurrencies from being considered a safe haven for hedging investments~\citep{ConlonT-2020a,ChoiS-2022a,James2022inf,WatorekM-2023a}.

Like any financial market where multiple assets are traded, the cryptocurrency market exhibits an internal structure~\citep{WatorekM-2021b}. From an investor’s perspective individual cryptocurrencies vary in significance with respect to factors such as trust, the governing consensus algorithm of the blockchain, transaction liquidity, and market capitalization. Moreover, investors in the cryptocurrency market tend to behave irrationally and are easily influenced by price changes, as well as concrete news items. This contrasts with traditional markets (e.g. stocks, commodities and currencies) where the dominance of professional investors exists. This results in more frequent herding behaviour among cryptocurrency price changes~\cite{nguyen2025herding}. Beyond this, groups of cryptocurrencies can be distinguished with different levels of internal coupling~\citep{wu2018classification,Duan2023,nguyen2023volatility,Alves2025,Bhattacherjee2025,Bouri2025,Zhou2025}. If this coupling is measured using correlation metrics between observables representing cryptocurrencies, such as returns, volatility, or transaction volume, it is possible to identify sectors within the market composed of cryptocurrencies with either similar characteristics or those treated similarly by investors. By using these correlation measures to represent the market as a network, a hierarchical structure emerges, where some cryptocurrencies play a central role in the network while others are secondary, tertiary, or peripheral. The most significant hubs that have been consistently identified across different studies are the most capitalized cryptocurrencies: BTC and ETH. The market sectors become network clusters and they are typically related to some secondary assets that play the role of the sectorial hubs~\citep{StosicD-2018a,ZiebaD-2019a,BriolaA-2022a,JamesN-2022a}. This structure is not stable but rather dynamic and evolves over time, however~\citep{SongJY-2019a,DrozdzS-2020a,KwapienJ-2021a,JamesN-2022b,JingR-2023a,Jin2025}.

Cross-correlation networks representing financial markets typically consist of a large number $N$ of nodes, each corresponding to a single asset, and an even greater number of weighted edges representing connections between individual asset pairs. In such cases, analyzing and especially visualizing the entire network quickly becomes cumbersome and unreadable. To address this issue, filters are applied to the complete network, removing insignificant or less relevant edges while retaining those that are crucial to the structure. There are many different types of such filters, varying in the number of nodes and edges they preserve and, thus, in the amount of information they retain. Among these, one can list a minimal spanning tree (MST)~\citep{MantegnaRN-1999a,OnnelaJP-2004a,KwapienJ-2009a}, a planar maximally filtered graph (PMFG)~\citep{TumminelloM-2005a,EryigitM-2009a,HongMY-2022a}, a triangulated maximally filtered graph (TMFG)~\citep{MassaraGT-2016a,MillingtonT-2022a,BriolaA-2022a}, and a correlation-threshold graph~\citep{BoginskiV-2005a,HuangWQ-2009a,TseCK-2010a}. Moreover, in addition to traditional correlation-based methods, more sophisticated measures based on distances between structural breaks~\citep{James2023e} and change points~\citep{James2020d} may also be considered as a base to network construction. However, it should be remembered that in order to be used here, these measures must satisfy the known three mathematical conditions of a metric, which may not always be possible. The measure based on the $\rho_q$ coefficient satisfies these conditions~\citep{KwapienJ-2015a,KwapienJ-2017a}.

Due to its structure and the small number of edges, the most commonly used filter is the minimum spanning tree (MST)~\citep{MantegnaRN-1999a}. It is a subnetwork of the complete network, consisting of $N$ nodes and $E = N-1$ edges, where the sum of the weights of all its edges is minimized. MST is constructed using either Prim’s algorithm~\citep{PrimRC-1957a} or Kruskal’s algorithm~\citep{KruskalJB-1956a}. The former performs better in dense networks ($E \sim N^2$) due to its $O(E + N \log N)$ complexity, while the latter is more efficient for sparse networks ($E \sim N$), with a complexity of $O(E \log N)$. The filter based on MST has been successfully used to analyze correlation structure in financial markets, such as stock markets~\citep{VandewalleN-2001a,MaslovS-2001a,OnnelaJP-2004a,CoelhoR-2007a,DjauhariMA-2012a,KwapienJ-2012a,JunD-2024a,WILINSKI2013,Miskiewicz2022}, the foreign exchange market~\citep{McDonaldM-2005a,GorskiAZ-2008a,KwapienJ-2009a,WangGJ-2012a,gkebarowski2019detecting,LiB-2020a,ZhangD-2022a}, commodity markets~\citep{SieczkaP-2009a,MatesanzD-2014a,MaYR-2021a,MagnerNS-2023a}, as well as the cryptocurrency~\citep{DurchevaM-2019a,JaureguizarFC-2018a,StosicD-2018a,ZiebaD-2019a,SongJY-2019a,PolovnikovK-2020a,Jaroonchokanan2025} and NFT markets~\citep{DeCollibusFM-2024a,WatorekM-2024a}. As regards the cryptocurrency market, listed in what follows are a few examples of such MST-based studies.

By examining the relationships among 100 cryptocurrencies in the years 2018-2019, expressed through cross-correlation coefficients and a measure of dissimilarity between periodograms for returns and volatility, it was demonstrated that the cryptocurrency market, represented by MST trees, has a hierarchical structure with a high degree of centralization, where the largest-capitalization coins were found to acts as hubs~\citep{DurchevaM-2019a}. A similarly sized group of 119 cryptocurrencies represented by daily data from the years 2016-2018 also exhibited a hierarchical structure but with significantly lower centralization~\citep{StosicD-2018a}. Based on a small set of 16 cryptocurrencies and the Pearson cross-correlation coefficients, a centralized market structure was reported for daily data from 2017-2018, with ETH as the dominant node~\citep{JaureguizarFC-2018a}. This transient dominant position of ETH as the network center was not an isolated event, as the situation repeated itself at the beginning of 2019~\citep{DrozdzS-2020a} and at the turn of 2020-2021~\citep{KwapienJ-2021a}. A study of cross-correlations within a set of 78 cryptocurrencies from 2015-2018 showed a developed sectoral structure already at that time. Furthermore, it was observed that BTC was relatively insensitive to external shocks and had little impact on the evolution of other cryptocurrencies. More influential assets were dogecoin (DOGE) and litecoin (LTC)~\citep{ZiebaD-2019a}.

Based on daily data from the years 2017-2018, a core-periphery structure in a network of 157 cryptocurrencies was identified, where the most capitalized coins were shifted to the periphery, while some less capitalized ones formed the center; this structure was explained through the trading properties of each coin~\citep{PolovnikovK-2020a}. In an analysis of high-frequency data for 76 cryptocurrencies collected over several months at the turn of 2017-2018, a clear dominance of BTC and ETH hubs was observed, which masked more subtle relationships among the remaining cryptocurrencies. In order to counter this effect, MST trees based on residual data after filtering out the influence of these two dominant assets were subsequently constructed. The residual structure had a strongly sectoral form with six distinguishable sectors, some of which were relatively stable and invariant to regulatory changes affecting the market during the studied period, while others had a more ephemeral nature~\citep{SongJY-2019a}. In another study, daily data for 136 cryptocurrencies was examined by using various correlation measures and constructing the related MSTs. The results demonstrated differences in the evolution of market structure depending on the correlation measure used. An increasing market correlation with rising market capitalization for some measures was reported, while other measures exhibited significantly greater randomness~\citep{Jaroonchokanan2025}. Daily price variations of a large set of 1000 cryptocurrencies with the largest capitalization were analyzed in order to investigate the reliability of constructing optimal portfolios based on cross-correlations among these cryptoassets. However, it was impossible to develop a profitable long-term investment by using this approach, because of a high degree of instability of the market cross-correlation structure, which required rebuilding the portfolios daily~\citep{JingR-2023a}. As the final example, a growing centralization of tokens (cryptocurrencies, DeFi's, and NFTs) on the Ethereum platform in agreement with the rich-get-richer paradigm was found in yet another recent study using the MST filtering~\citep{DeCollibusFM-2024a}.

The standard MST approach, which applies the Pearson cross-correlation coefficient as a measure of bivariate interdependence between time series, proves problematic for nonstationary data as the results can be unreliable~\citep{KwapienJ-2017a}. Trends represent a prominent source of statistical nonstationarity if present in the time series. Thus, they need to be eliminated first. The most popular framework to deal with trends is detrended fluctuation analysis (DFA)~\citep{PengCK-1994a} together with its multiscale generalization - multifractal detrended fluctuation analysis~\citep{KantelhardtJ-2002a}. Together with its bivariate variant - multifractal detrended cross-correlation analysis (MFCCA)~\citep{PodobnikB-2008a,ZhouWX-2008a,OswiecimkaP-2014a} - these two latter methods can be combined in order to define a $q$-dependent detrended cross-correlation coefficient $\rho_q$, which may be used as a direct counterpart of the Pearson coefficient for nonstationary data $\rho_{\rm DCCA}$ if $q=2$. However, since $q \in \mathbb{R}$, the new measure offers much more than that. By adjusting the value of the parameter $q$, one can select the magnitude of local variances of the detrended signals and focus only on their selected parts. In this way, the cross-correlation structure of the analyzed data may be broken down to a specific range of amplitudes only~\citep{KwapienJ-2015a}. With the use of $\rho_q$, one may proceed to multivariate sets, construct the corresponding $q$-dependent cross-correlation matrix, and finally, arrive at the definition of the $q$-dependent detrended MST constructed in the same way as the regular MSTs but here based on the $q$-order detrended cross-correlations~\citep{WangGJ-2013a,KwapienJ-2017a}.

The $q$-dependent detrended minimum spanning trees ($q$MSTs) were proposed as a tool for visualizing a selective cross-correlation structure of multivariate non-stationary time series filtered based on fluctuation magnitude~\citep{KwapienJ-2017a}. It was shown that, by applying the $q$-dependent detrended cross-correlation coefficient $\rho_q$ to time series representing stock returns of the largest US companies, it was possible to extract genuine information on the cross-correlation structure of the US market that could not be obtained based on the detrended cross-correlation coefficient $\rho_{\rm DCCA}$. It was also shown that the $q$MST topology could differ substantially between the graphs constructed for different values of $q$. For small and medium returns ($q \leqslant 2$), the respective $q$MSTs had a centralized structure, while large returns ($q > 2$) developed trees the more dispersed, the larger return amplitude was considered~\citep{KwapienJ-2017a}.

There are a few studies available in literature, in which the idea of $q$MSTs have been successfully exploited. Zhao et al.~\citep{ZhaoL-2018a} analyzed the cross-correlation structure of price returns for a set of 401 S\&P500-constituent stocks and Lin et al.~\citep{LinM-2018a} studied cross-correlations between 37 world stock-market indices representing major economies by focusing on fluctuations of different magnitude. In these works, planar maximally filtered graphs (PMFGs~\citep{TumminelloM-2005a}) based on the $q$-dependent detrended cross-correlation coefficient $\rho_q$ were constructed. Like their standard counterparts do with MSTs, these $q$PMFGs contain $q$MSTs as their subnetworks created in the initial step of the construction algorithm. Quite surprisingly, in both the S\&P500 stocks and the world indices, the obtained results showed that small returns are cross-correlated more strongly than the large ones, a result that has seldom been reported in literature. However, different subsets of stocks and different subsets of indices were correlated in each case. From the node degree perspective, the returns of small or medium amplitude formed networks with significant heterogeneity, while large returns revealed networks that were much more homogeneous. Application of the obtained $q$PMFGs to optimal portfolio selection under the mean-variance framework by using centrality measures as the selection metric showed that portfolios based on peripheral stocks outperform the ones based on central stocks with $q=2$ as the best choice under the condition of the largest difference in network topology. However, the same analysis carried out under the expected shortfall framework pointed out to the values $2 \leqslant q \leqslant 6$ instead~\citep{ZhaoL-2018a}.

$q$MST trees themselves were then applied to study the structure of cross-correlations in the cryptocurrency market~\citep{KwapienJ-2021a}. High-frequency data for the 80 largest-capitalization cryptocurrencies traded on the Binance exchange were analyzed using a rolling window to determine changes in the correlation structure over time during the years 2020-2021. This structure underwent significant changes during that period, becoming increasingly centralized. For short time scales on the order of minutes, the $q$MST tree structure was found to have a single star-like shape centered on either BTC or ETH. These shifts between the two main cryptocurrencies were rare for these time scales, and the networks remained relatively stable. However, the situation was entirely different for longer time scales, on the order of several hours, where the central hub frequently changed, switching among the most liquid cryptocurrencies. In addition to BTC and ETH, assets such as ontology (ONT), tron (TRX), and FTX token (FTX) also played central roles. The results also showed that while the cryptocurrency market was relatively independent of other financial markets at the onset of the pandemic, as markets ed to the pandemic and its impact decreased, inter-market cross-correlations became strong again.

In this work, a set of time series representing price returns of highly capitalized cryptocurrencies is analyzed by means of the coefficient $\rho_q$ and the $q$MST graphs in order to investigate the temporal evolution of the cross-correlation structure of the cryptocurrency market. The paper is organized as follows: in Sect.~\ref{data_meth}, the essential information regarding the MFDFA/MFCCA methodology, the coefficient $\rho_q$, and the $q$MST graph is provided together with a brief description of the datasets used. In Sect.~\ref{results} the results are reported and discussed, while conclusions and future research perspectives are presented in Sect.~\ref{concl}.

\section{Data and Methods}
\label{data_meth}

\subsection{Multifractal detrended cross-correlation analysis}
\label{mfcca}

It happens frequently that empirical time series recorded from observables related to natural complex systems reveal a multifractal organization of fluctuations~\citep{KwapienJ-2012a,DrozdzS-2015a}. Identification and quantification of such an organization in time series requires a properly designed methodology that is able to grasp the genuine effects and neglect spurious ones. It has already been demonstrated that one of the possible choices in this respect is MFDFA/MFDCCA, a methodology that proved to be effective and reliable~\citep{KantelhardtJ-2002a,OswiecimkaP-2006a,ZhouWX-2008a,OswiecimkaP-2014a,JiangZQ-2019a}. It is based on observing scaling properties of the moments of time series that have been detrended. This methodology can be described as follows. Let one consider two time series $U=\{u(i)\}_{i=1}^T$ and $V=\{v(i)\}_{i=1}^T$ of length $T \gg 1$ that are sampled at the same time instants $i$. First, both time series are integrated to form their profiles $\tilde{U}$ and $\tilde{V}$
\begin{equation}
\tilde{u}(i)= \sum_{j=1}^i u(j), \quad \tilde{v}(i)= \sum_{j=1}^i v(j),
\label{eq::mfdcca.profiles}
\end{equation}
respectively. These time series can be divided into $M_s$ segments of length $s$ each starting from both their beginnings and their ends in order not to neglect any data points, so there are $2M_s$ segments total. Next, a detrending procedure is applied, in which a polynomial trend $P_{\nu}^{(m)}$ of order $m$ is subtracted from $\tilde{U}$ and $\tilde{V}$ in each segment $\nu$ independently:
\begin{eqnarray}
\nonumber
x(\nu s + k) = \tilde{u}(\nu s + k) - P_{\textrm{X},\nu}^{(m)}(k),\\
y(\nu s + k) = \tilde{v}(\nu s + k) - P_{\textrm{Y},\nu}^{(m)}(k),
\label{eq::mfdcca.detrending}
\end{eqnarray}
where $k=1,...,s$ and $\nu=0,...,2M_s-1$. In the subsequent step, segment-wise covariance $f_{\textrm{XY}}^2$ and variances $f_{\textrm{XX}}^2$, $f_{\textrm{YY}}^2$ are calculated
\begin{eqnarray}
\nonumber
f_{\textrm{XY}}^2(s,\nu) = {1 \over s} \sum_{k=1}^s x(\nu s + k) y(\nu s + k),\\
f_{\textrm{XX}}^2(s,\nu) = {1 \over s} \sum_{k=1}^s x^2(\nu s + k),\label{eq::mfdcca.covariances}\\
f_{\textrm{YY}}^2(s,\nu) = {1 \over s} \sum_{k=1}^s y^2(\nu s + k).
\nonumber
\end{eqnarray}
Then a family of bivariate fluctuation functions $F_{\textrm{XY}}^q$ and univariate ones $F_{\textrm{XX}}^q$, $F_{\textrm{YY}}^q$ is obtained by raising, respectively, $f_{\textrm{XY}}^2$, $f_{\textrm{XX}}^2$, and $f_{\textrm{YY}}^2$ to a real power $q$ and taking averages over the segments:
\begin{eqnarray}
\nonumber
F_{\textrm{XY}}^q(s) = \bigg\{ {1 \over 2 M_s} \sum_{\nu=0}^{2 M_s - 1} {\rm sign} \left[f_{\textrm{XY}}^2(s,\nu)\right] |f_{\textrm{XY}}^2(s,\nu)|^{q/2} \bigg\}^{1/q}, \\
F_{\textrm{XX}}^q(s) = \bigg\{ {1 \over 2 M_s} \sum_{\nu=0}^{2 M_s - 1} \left[f_{\textrm{XX}}^2(s,\nu)\right]^{q/2} \bigg\}^{1/q},
\label{eq::mfdcca.fluctuation-functions}\\
\nonumber
F_{\textrm{YY}}^q(s) = \bigg\{ {1 \over 2 M_s} \sum_{\nu=0}^{2 M_s - 1} \left[f_{\textrm{YY}}^2(s,\nu)\right]^{q/2} \bigg\}^{1/q}.
\end{eqnarray}
The sign function in the first formula in Eq.~(\ref{eq::mfdcca.fluctuation-functions}) has been introduced in order to guarantee that the fluctuation functions remain real for any choice of the parameter $q$, but also for consistency of the results~\citep{OswiecimkaP-2014a}. 

All the steps described so far are repeated for different values of the segment length $s$. If the time series $\textrm{X}$, $\textrm{Y}$ are fractal, the univariate fluctuation functions $F_{\textrm{XX}}^q$, $F_{\textrm{YY}}^q$ show a power-law dependence on $s$~\citep{KantelhardtJ-2002a}:
\begin{eqnarray}
F_{\textrm{XX}}^q(s) \sim s^{h_{\textrm{X}}(q)}, \quad F_{\textrm{YY}}^q(s) \sim s^{h_{\textrm{Y}}(q)}.
\label{eq::mfdcca.univariate.scaling}
\end{eqnarray}
The exponents $h_{\textrm{X}}(q)$ and $h_{\textrm{Y}}(q)$ are non-increasing functions of $q$ and are called the generalized Hurst exponents, because they coincide with the Hurst exponent $H$ for $q=2$. Based on their behaviour, the following two cases can be distinguished: a monofractal scaling if $h_{\cdot}(q)=\textrm{const}$ and a multifractal one if $h_{\cdot}(q)$ is monotonously decreasing in $q$. If, in addition, the bivariate fluctuation function $F_{\textrm{XY}}^q$ defined by the following formula
\begin{equation}
F_{\textrm{XY}}^q(s) \sim s^{\lambda_{\textrm{XY}}(q)},
\end{equation}
shows scaling, the two time series X, Y are said to be monofractally cross-correlated if $\lambda_{\textrm{XY}}(q)=\textrm{const}$ or multifractally cross-correlated otherwise. The parameter $q \in \mathbb{R}$ plays an important role in allowing for the extraction of fluctuations within the amplitude range of interest by selectively amplifying them and attenuating those in other amplitude ranges. The standard case corresponding to equally weighted fluctuations is obtained for $q=2$.

\subsection{$q$-dependent detrended minimum spanning trees}

The univariate and bivariate fluctuation functions can be used to define the $q$th-order detrended cross-correlation coefficient $\rho_q(s)$
\begin{equation}
\rho_q(s) = {F_{\textrm{XY}}^q(s) \over \sqrt{ F_{\textrm{XX}}^q(s) F_{\textrm{YY}}^q(s) }}.
\label{eq::rhoq}
\end{equation}
introduced in~\citep{KwapienJ-2015a} as a generalization of the detrended cross-correlation coefficient $\rho_{\rm DCCA}$~\citep{ZebendeGF-2011a}. The role of the parameter $q$ is similar to the one it plays in the case of the fluctuation functions. However, while in principle $q \in \mathbb{R}$ also in this case, there are some subtleties regarding the behaviour of the coefficient for different ranges of $q$. For $q \geqslant 0$, values of $\rho_q(s)$ satisfy the condition $-1 \le \rho_q \le 1$, which is not the case for $q < 0$, where the coefficient may assume values outside this range (for more information, see~\citep{KwapienJ-2015a}). However, consideration of such a situation is beyond the scope of the present study, in which we restrict our analysis to $q \geqslant 0$. It is important to note that, being a function of scale, the coefficient $\rho_q(s)$ does not require the fluctuation functions used in its calculation to exhibit a power-law dependence, which makes it a robust tool that can also be used for non-fractal time series. The formula (\ref{eq::rhoq}) implies invariance of $\rho_q(s)$ under a swap of time series $\textrm{X} \leftrightarrow \textrm{Y}$.

In a multivariate case, when $N$ parallel time series are of interest, in order to obtain a complete correlation map, one has to compute $N(N-1)/2$ values of $\rho_q(s)$ for each considered time scale $s$. It is thus convenient to arrange these values in an $N \times N$ matrix ${\bf C}(q,s)$ with elements $C_{ij}(q,s) \equiv \rho_q^{ij}(s)$, which may be considered as a $q$-dependent detrended cross-correlation matrix ($i,j=1,...,N$). It can be diagonalized and its eigenvalues $\lambda_i$ and eigenvectors ${\bf v}_i$ can be obtained by using the formula
\begin{equation}
{\bf C}(q,s) {\bf v}_i(q,s) = \lambda_i(q,s) {\bf v}_i(q,s).
\label{eq::eigenspectrum}
\end{equation}

Due to the fact that $\rho_q(s)$ cannot be used as a metric (similar to the Pearson correlation coefficient, $\rho_q$ doesn't satisfy the triangle inequality condition for time series triples), one has to redefine the matrix elements $C_{ij}(q,s)$ to partially address this issue:
\begin{equation}
D_{ij}(q,s) = \sqrt {2 \left[ 1-C_{ij}(q,s) \right]},
\label{eq::distance.matrix}
\end{equation}
where $D_{ij}(q,s)$ are the elements of a distance matrix ${\bf D}(q,s)$. Now these elements satisfy the triangle inequality if $q \geqslant 1$ (see~\citep{KwapienJ-2017a} for a related discussion).

Based on the matrix ${\bf D}(q,s)$, one may construct a weighted, undirected network $\mathcal{N}$ consisting of $N$ nodes, with each node representing a time series under study. Then, by applying Kruskal's or Prim's algorithm~\citep{KruskalJB-1956a,PrimRC-1957a,MantegnaRN-1999a} to the elements $D_{ij}(q,s)$ for fixed $q$ and $s$, one can extract a subset of $\mathcal{N}$ with the same number of nodes and $N-1$ undirected edges that minimize the edge weight sum. This subset is called the $q$-dependent detrended minimum spanning tree ($q$MST) of $\mathcal{N}$~\citep{WangGJ-2013a,KwapienJ-2017a}. As being based on the coefficients $\rho_q(s)$, $q$MST is sensitive, by construction, to the segment-wise detrended covariances (Eq.~(\ref{eq::mfdcca.covariances})) of the considered multivariate time series. Therefore, one can focus on a particular range of covariances by amplifying the relative contribution of particular segments $\nu$ to $F_{\textrm{XY}}^q(s)$ and suppressing the relative contribution of the remaining ones. In consequence, a resulting $q$MST may reflect the multivariate structure of, for example, strong ($q > 2$) or moderate ($q < 2$) covariances rather than the overall average covariance structure ($q=2$). Although small covariances ($q < 0$) cannot be selected in this way due to the necessary condition $q \geqslant 1$ for $q$MST, this does not pose a problem because small covariances are likely statistically insignificant. In the empirical part of the paper, two representative values $q=1$ and $q=4$ will be considered. In the case of $q=1$, there is no additional relative amplification of fluctuations~\citep{KwapienJ-2015a}, thus the average fluctuations play the most significant role. The value of $q=4$ was chosen in order to amplify the effect of large fluctuations relative to the smaller ones. In the present case this value of $q$ also sets the upper limit allowing the convergence of moments~\citep{Embrechts1997}. For $q>4$, the moments may diverge due to the inverse cubic power-law (tail exponent of about 3)  governing the asymptotic distribution of large returns~\citep{gopikrishnan1998}, also in the cryptocurrency exchange rates case~\citep{WatorekM-2021a}.

\subsection{Data specification}
\label{data}
The data set comprises $N=140$ exchange rates of the most traded cryptocurrencies expressed in USDT on the Binance exchange~\citep{Binance}, covering the period from January 1, 2021 to September 30, 2024 (the data are available in an open repository~\citep{DataBinance}). As Binance is the largest exchange with the highest volume value~\citep{coinmarket}, the data set includes all highly capitalized cryptocurrencies and thus it is representative of the entire cryptocurrency market. The stablecoins were excluded from the analysis because their volatility relative to USDT is minimal.

The time series were sampled at 1-minute frequency. The exchange rate time series were first transformed into logarithmic returns $R_i(t_m)=\ln p_i(t_{m+1})-\ln p_i(t_m)$, where $m=1,...,T-1$ and $i$ represents a specific cryptocurrency ticker. The complete list of the cryptocurrency tickers considered in this study, along with their respective sectors according to the CoinDesk classification~\citep{coindesk}, is provided in Appendix~\ref{secA1} in Tab.~\ref{tab::ticker_list}. Basic statistics for each of the exchange rate are also included: the average volume value $\langle V_{\Delta t} \rangle$, the average number of transactions $\langle N_{\Delta t} \rangle$, and the fraction of zero log-returns $\%0 R_{\Delta t}$ for $\Delta t$ = 1min. There are visible significant differences between cryptocurrencies in average volume and trading frequency in the data set considered. BTC has the first position in both statistics, and ETH is clearly 2nd, with a significant gap to the rest.

\begin{figure}[ht!]
\centering
\includegraphics[width=0.99\textwidth]{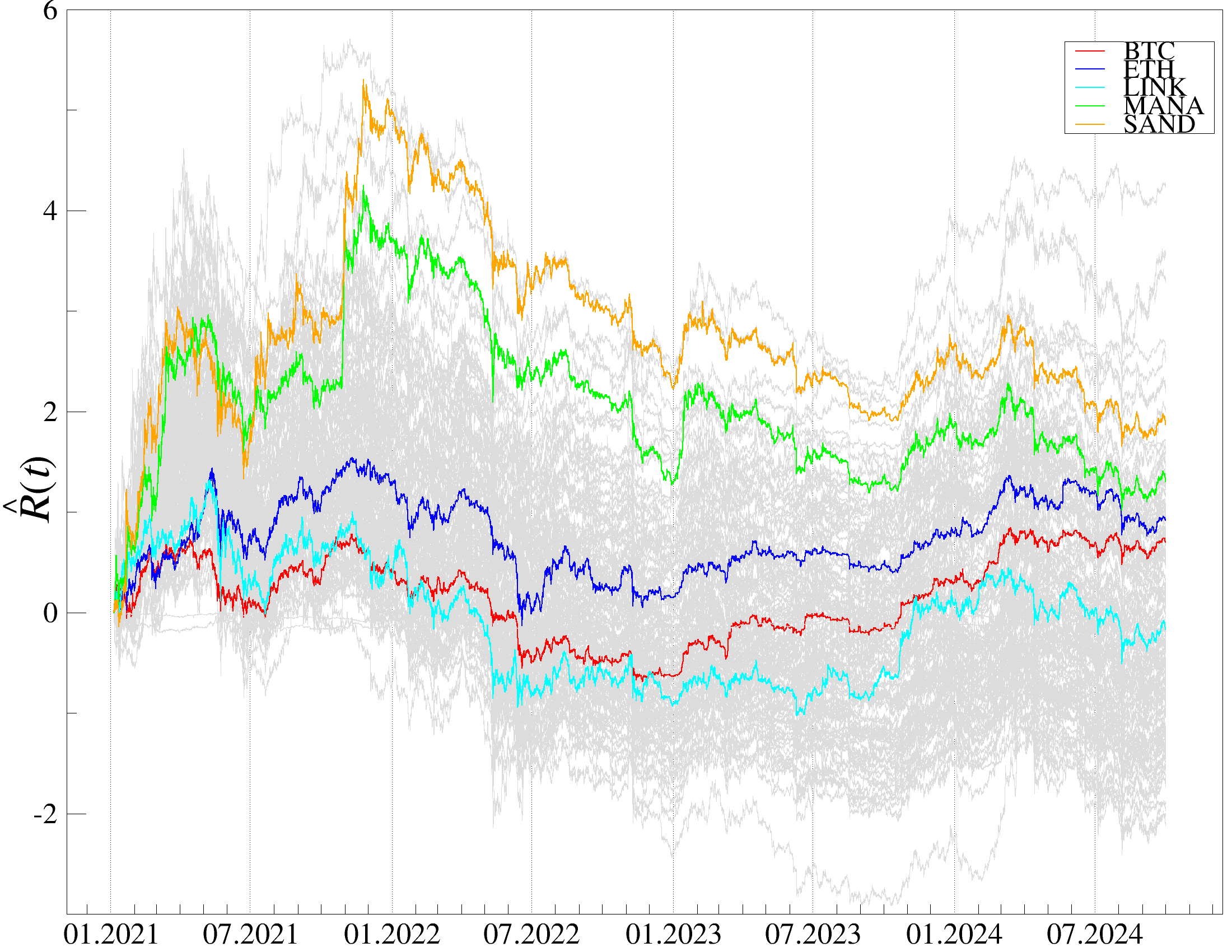}
\caption{Evolution of the cumulative log-returns $\hat{R}(t)$ of the 140 cryptocurrencies over the time period from Jan 1, 2021 to Sep 30, 2024. The colors of two of the most liquid cryptocurrencies and a few other distinguished ones are indicated explicitly. The bulk of the cryptocurrencies is shown in the background (grey lines).}
\label{fig::returns.integrated}
\end{figure}

The evolution of the cumulative logarithmic returns $\hat{R}(t_m)=\sum_{m=1}^T R(t_m)$ of the 140 cryptocurrencies over the analyzed time period is presented in Fig.~\ref{fig::returns.integrated}. Various phases of the market can be observed. The bull market in 2021, then the bear market in 2022, with the crash in May 2022. After reaching its bottom at the end of 2022, the market was in a slower growth phase until mid-2024, then moved sideways until the end of September 2024. Thus, the selected dataset allows for market analysis under various conditions.

% Figure 1

\section{Results}
\label{results}

\subsection{Changes in cross-correlations over time}
\label{sect::temporal.evolution}

To track the evolution of cross-correlations, a rolling window of 7 days - trading week (10,080 data points) was applied, with a daily step (1,440 data points) along the time series. In each rolling window, ${\bf C}(q,s)$ (Eq.~\ref{eq::eigenspectrum}) was calculated and then transformed into ${\bf D}(q,s)$ (Eq.~\ref{eq::distance.matrix}), from which the $q$MST graph was constructed. Then, the spectral and network characteristics were calculated for them. These include the largest eigenvalue $\lambda_1$, the squared expansion coefficients of the eigenvector ${v}^{2}_{1,j}$ associated with $\lambda_1$, the Shannon entropy of the squared eigenvector component, defined by $H({\bf v}^{2}_{1})= - \sum_{j=1}^N v^2_{1,j} \ln v^2_{1,j}$, node degree $k$, and average path length $\langle L \rangle = {1 \over N(N-1)} \sum_{i=1}^N \sum_{j=i+1}^N L_{ij}$, where $L_{ij}$ is the length of the path connecting nodes $i$ and $j$. 

\begin{figure}[ht!]
\centering
\includegraphics[width=0.99\textwidth]{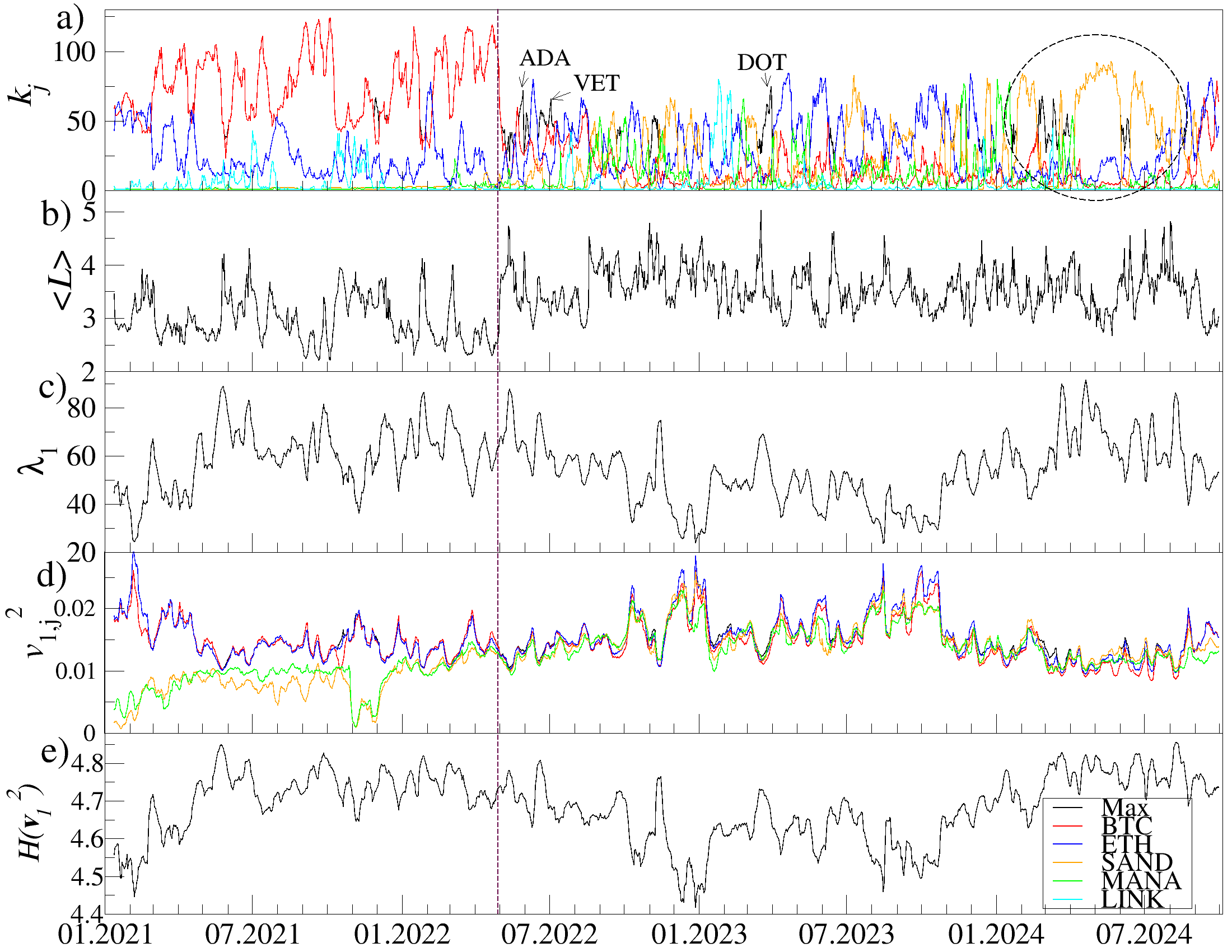}
\caption{Time evolution of the network characteristics of the $q$MSTs created from a distance matrix ${\bf D}(q=1,s=10)$: (a) node degree $k_j$ (cryptocurrencies that had the highest multiplicity in a given window were indicated), (b) average path length $\langle L \rangle$ and spectral characteristics of the $q$-dependent detrended correlation matrix ${\bf C}(q=1,s=10)$: (c) the largest eigenvalue $\lambda_1$, (d) the squared expansion coefficients of the eigenvector ${ v}^{2}_{1,j}$ associated with $\lambda_1$ for $j$=BTC, ETH, MANA, LINK, and SAND (e) the Shannon entropy H(${\bf v}^{2}_{1}$) of the squared eigenvector components. Rolling window of length 7 days shifted by 1 day was applied.}
\label{fig::rolling-windowq=1s=10}
\end{figure}

Fig.~\ref{fig::rolling-windowq=1s=10} presents the changes in selected characteristics over time for the correlation matrix ${\bf C}(q=1,s=10)$, corresponding to a scenario when the average fluctuations play the most significant role, at the shortest possible time scale $s=10$min selected due to the sufficient length of the segment in detrending procedure (Eq.~\ref{eq::mfdcca.detrending})~\citep{Oswiecimka2013}, with the sampling frequency of 1 min. A significant shift in network characteristics is evident starting from the rolling window ending at the end of April 2022 (as indicated by a dashed line in Fig.~\ref{fig::rolling-windowq=1s=10}). Before this date, the MST structure was more centralized, with BTC being the highest multiplicity node in most windows. This centralized structure corresponds to a low average path length $\langle L \rangle$. In contrast, since May 2022, the MST has become more decentralized with $\langle L \rangle$ almost always above 3 (Fig.~\ref{fig::rolling-windowq=1s=10}b) and the maximum node degree never exceeding 90 (Fig.~\ref{fig::rolling-windowq=1s=10}a). Moreover, during this later period, BTC loses its dominant role, and various cryptocurrencies, such as ADA, DOT, ETH, LINK, MANA, SAND, and VET, emerge as the largest-multiplicity nodes.

% Figure 2

\begin{figure}[ht!]
\centering
\includegraphics[width=0.49\textwidth]{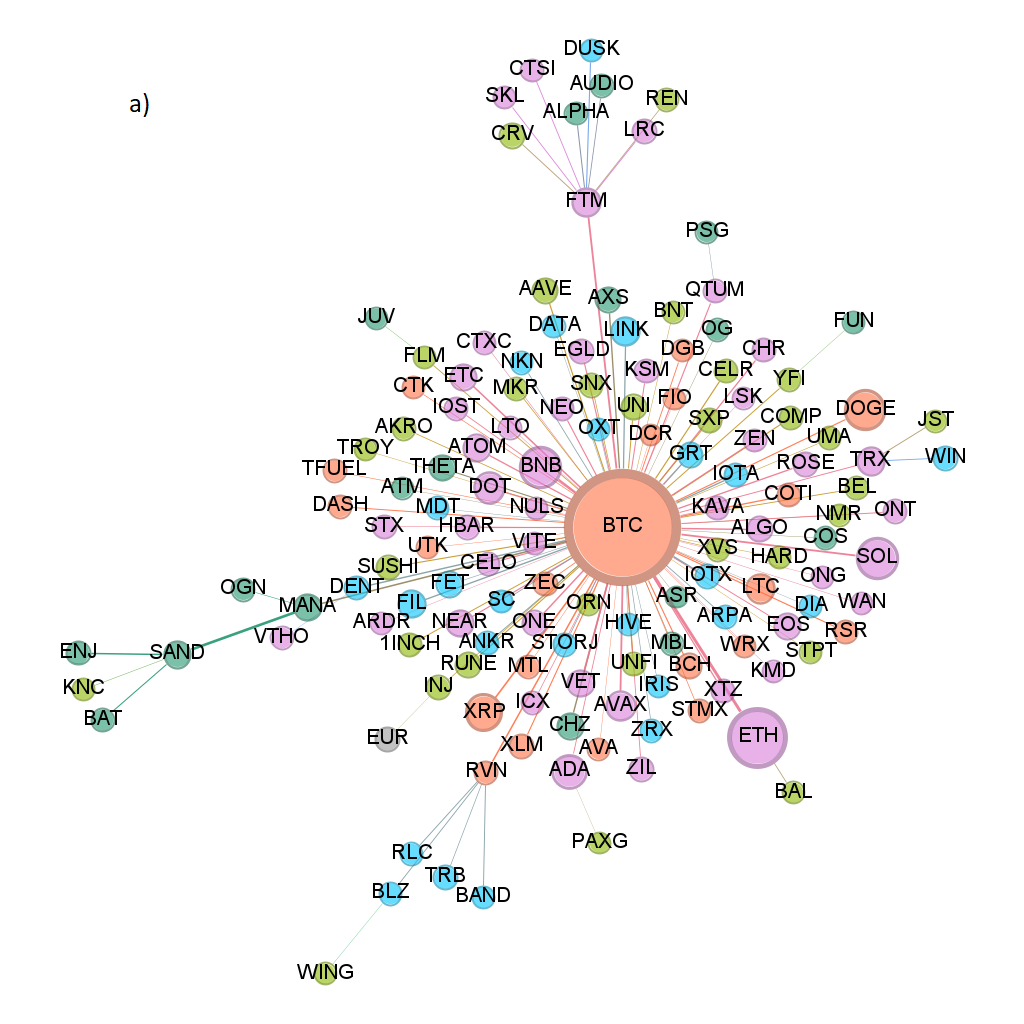}
\includegraphics[width=0.49\textwidth]{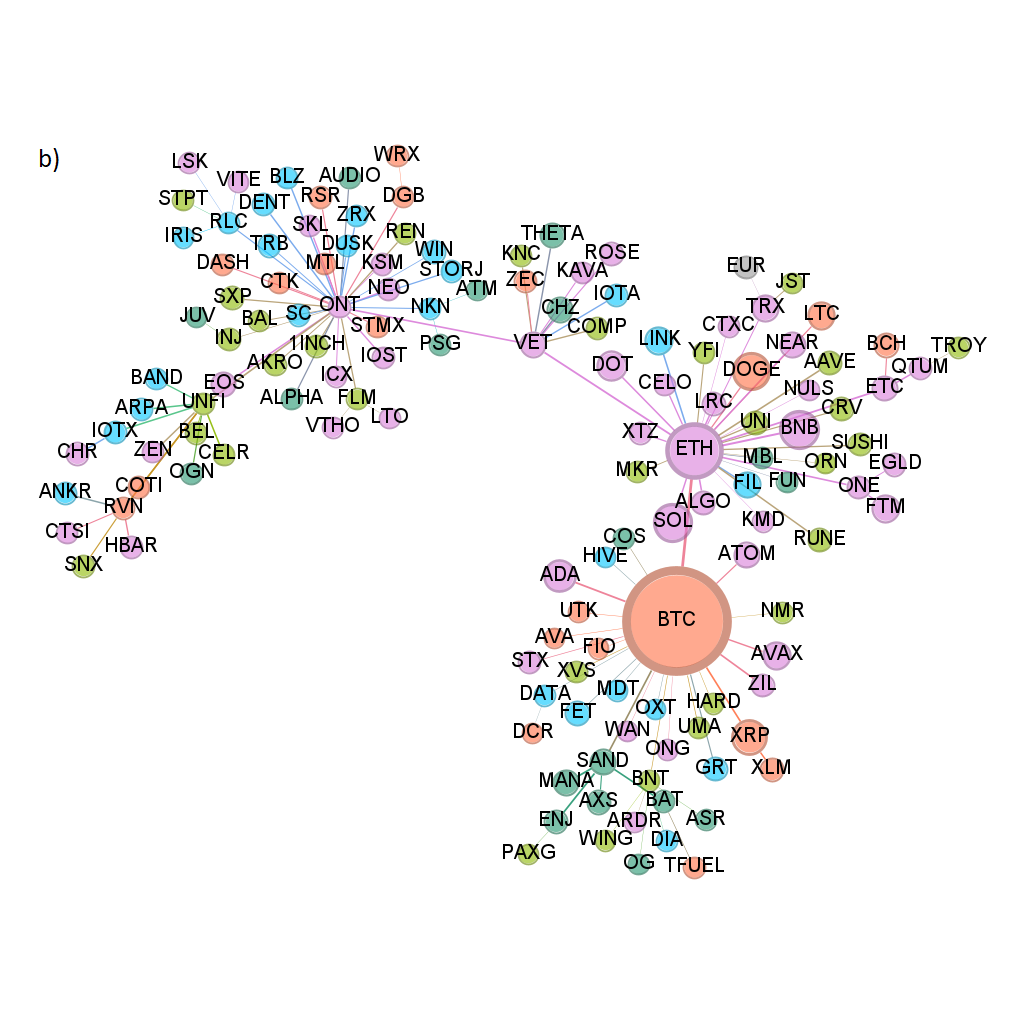}
\caption{Sample $q$MSTs calculated for ($q=1$ and $s=10$) in the rolling windows ending on: (a) Apr 25, 2022 and (b) May 18, 2022. Node size is proportional to the average volume in the analyzed period, while edge thickness reflects the strength of the cross-correlations. Colors represent market sectors after Digital Asset Classification Standard (DACS), created by CoinDesk~\citep{coindesk}: currency (orange), smart contract platform (violet), computing (cyan), DeFi (green), and culture \& entertainment (dark green).}
\label{fig::MST1q}
\end{figure}

The change in the structure of the network in May 2022 is clearly visible in Fig.~\ref{fig::MST1q} where two sample $q$MSTs calculated within rolling windows at their respective endpoints are shown: (a) 25 April-2022 and (b) 18 May-2022. Each node represents a specific cryptocurrency, while each weighted edge indicates the metric distance between a pair of cryptocurrencies. In Fig.~\ref{fig::MST1q}a, the network structure is highly centralized, with BTC serving as the clearly largest multiplicity node with $k=112$. In Fig.~\ref{fig::MST1q}b, the MST structure is in the process of changing to decentralized and the largest multiplicity node is ONT with $k=33$. 

% Figure 3

The regime change in May 2022 is also evident in the spectral characteristics of the correlation matrix. Before May 2022, cross-correlations were generally stronger as indicated by larger $\lambda_1$ values (Fig.~\ref{fig::rolling-windowq=1s=10}c), which represent the market factor correlations. The diminishing dominance of BTC in the MST structure is further reflected in the reduced value of its expansion coefficient of the eigenvector ${\bf v}_{1,j}$ associated with $\lambda_1$ (Fig.~\ref{fig::rolling-windowq=1s=10}d). Changes in the largest eigenvalue over time correspond to changes in the Shanon entropy of the eigenvector components associated with $\lambda_1$ (Fig.~\ref{fig::rolling-windowq=1s=10}e). Stronger correlations (larger $\lambda_1$) indicate a more uniform share of individual cryptocurrencies in the correlations. Weaker correlation (lower $\lambda_1$) denote greater differentiation between the share of individual cryptocurrencies in the eigenvector. The observed shift in the cross-correlation structure at the end of April 2022 may be linked to the cryptocurrency market crash that developed at that time and accelerated in May 2022 following the collapse of Terra/Luna system~\cite{Briola2023}. 

Another interesting period can be observed from February to August 2024, where in most rolling windows, the cryptocurrency SAND, related to metaverse Sandbox became the largest multiplicity node, with $k=90$ in some windows. This is in conjunction with an increase of $\lambda_1$ entropy values to the 2022 level. Additionally, the notable position of SAND is confirmed, as it exhibits the highest expansion coefficient values in some rolling windows at that period. At the same time, it should be noted that the highest SAND node multiplicity did not match the BTC node degree from 2021 and 2022.

\subsection{Differences in the organization of cross-correlations at various fluctuation amplitudes}
\label{sect::differences}

% Figure 4

\begin{figure}[ht!]
\centering
\includegraphics[width=0.99\textwidth]{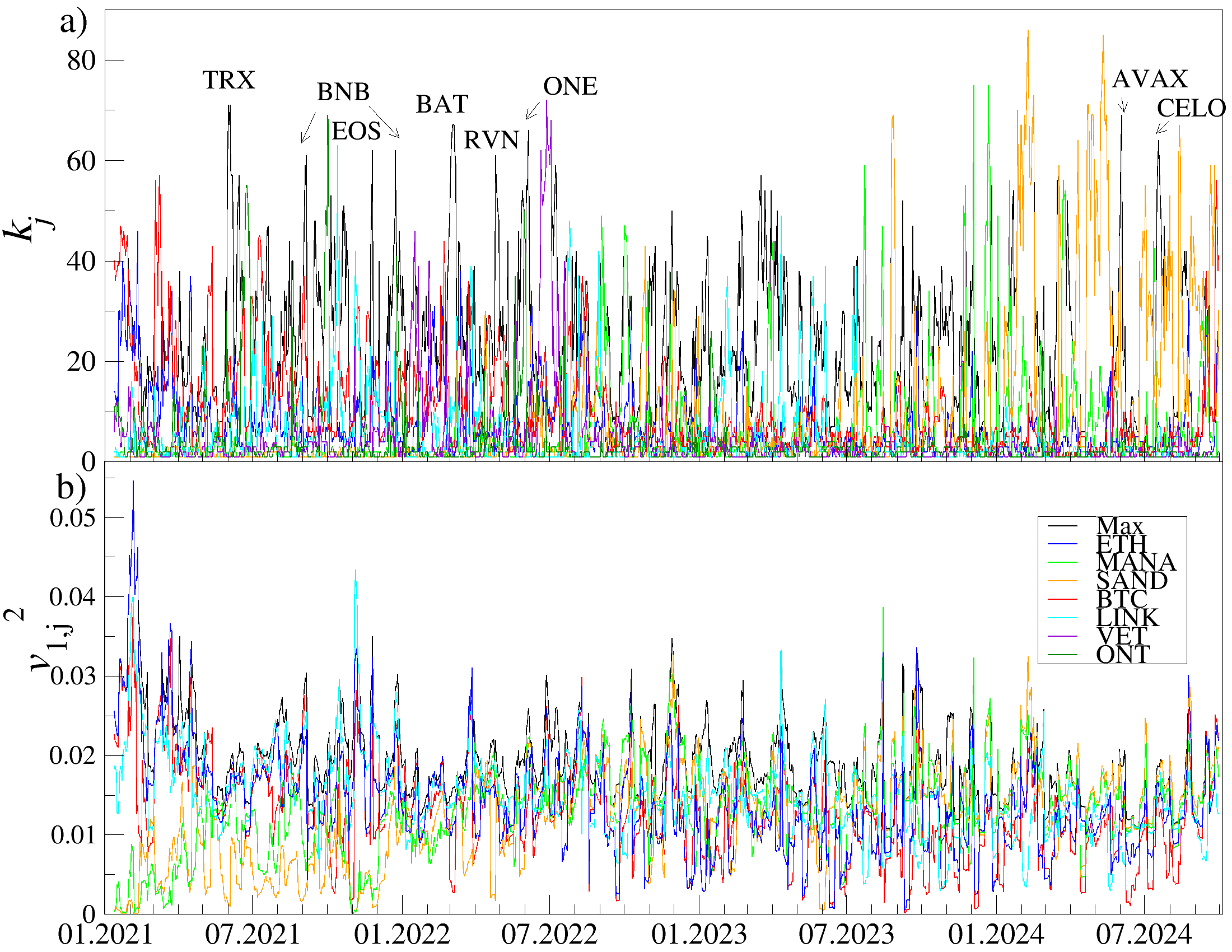}
\caption{Time evolution of the network characteristics of the $q$MSTs created from a distance matrix ${\bf D}(q=4,s=10)$: (a) node degree $k_j$ (cryptocurrencies that had the largest multiplicity in a given window were indicated) and spectral characteristics of the $q$-dependent detrended correlation matrix ${\bf C}(q=4,s=10)$: (b) the squared expansion coefficients of the eigenvector ${\bf v}^{2}_{1,j}$ associated with $\lambda_1$ for $j$=BTC, ETH, SAND, MANA, and LINK.}
\label{fig::rolling-windowq=4s=10}
\end{figure}

\begin{figure}[ht!]
\centering
\includegraphics[width=0.99\textwidth]{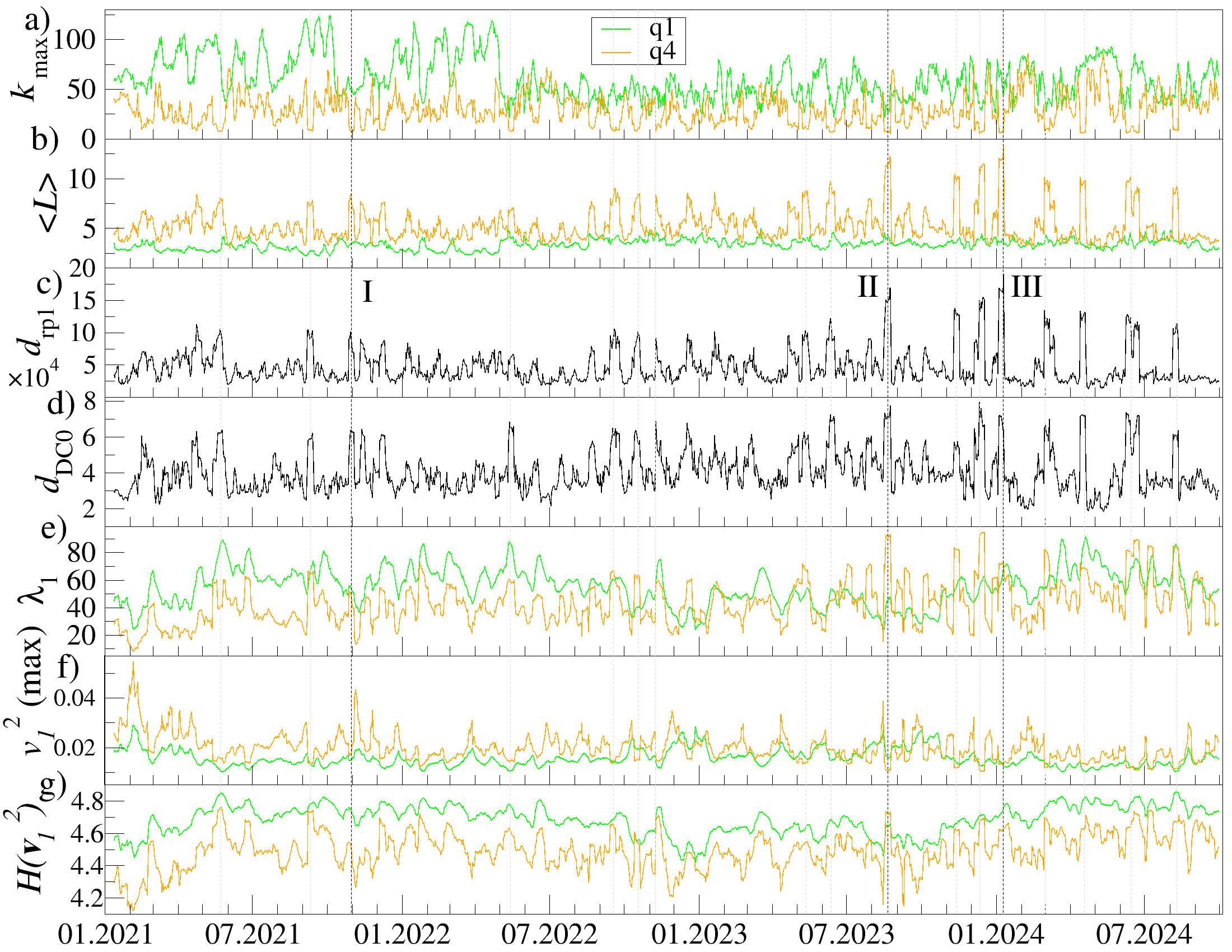}
\caption{Time evolution of the network characteristics of the $q$MSTs created from a distance matrix ${\bf D}(q=1,s=10)$ and ${\bf D}(q=4,s=10)$: (a) max node degree $k_{\textrm{max}}$, (b) average path length $\langle L \rangle$, (c) $d_{\textrm{rp}1}$, and (d) $d_{\textrm{DC}0}$ between $q=1$ and $q=4$ MST. The spectral characteristics of the $q$-dependent detrended correlation matrix ${\bf C}(q=1,s=10)$ and ${\bf C}(q=4,s=10)$: (e) the largest eigenvalue $\lambda_1$, (f) the highest squared expansion coefficients of the eigenvector ${\bf v}^{2}_{1,\textrm{max}}$ associated with $\lambda_1$, and (g) the Shannon entropy H(${\bf v}^{2}_{1}$) of the squared eigenvector components. Rolling window of length 7 days shifted by 1 day was applied. Periods with large intraday drops are marked with Roman numerals and dotted lines.}
\label{fig::rolling-windowq1_q4_s=10}
\end{figure}

The structure of the $q$MST graphs looks different if the correlations between the largest fluctuations are amplified ($q=4$). Here, the largest node degree changes significantly more often and the MST structure is less stable than in the case of $q=1$. In Fig.~\ref{fig::rolling-windowq=4s=10} there is no analogous period of BTC dominance corresponding to Fig.~\ref{fig::rolling-windowq=1s=10}. In addition, the network structure is more decentralized. This manifests itself in correlations and network characteristics presented in Fig.~\ref{fig::rolling-windowq1_q4_s=10}. The largest node degree is significantly lower and $\langle L \rangle$ is larger for $q=4$ (Fig.~\ref{fig::rolling-windowq1_q4_s=10}a and Fig.~\ref{fig::rolling-windowq1_q4_s=10}b) than for $q=1$. The differences are also visible in the spectral characteristics of the correlation matrices. The correlations measured using the largest eigenvalue $\lambda_1$ are larger for $q=1$ than for $q=4$ for most of the period (Fig.~\ref{fig::rolling-windowq1_q4_s=10}e). The largest expansion coefficient in the eigenvector associated with $\lambda_1$ is larger for $q=4$, which indicates greater differentiation among the eigenvector components (Fig.~\ref{fig::rolling-windowq1_q4_s=10}f) and, thus, lower Shannon entropy of the eigenvector components associated with $\lambda_1$ for $q=4$ (Fig.~\ref{fig::rolling-windowq1_q4_s=10}g). 

Despite the fact that the described dependencies occur in the vast majority of windows in Fig.~\ref{fig::rolling-windowq1_q4_s=10}, there are a few exceptions when $\lambda_1$ values are larger for $q=4$ than for $q=1$. In these windows, the structure of the eigenvector expansion coefficients associated with $\lambda_1$ for $q=4$ is homogeneous and the entropy is higher than for $q=1$. This also affects the network structure that is fully decentralized for $q=4$, which manifests itself through $\langle L \rangle > 10$. 

\begin{figure}[ht!]
\centering
\includegraphics[width=0.49\textwidth]{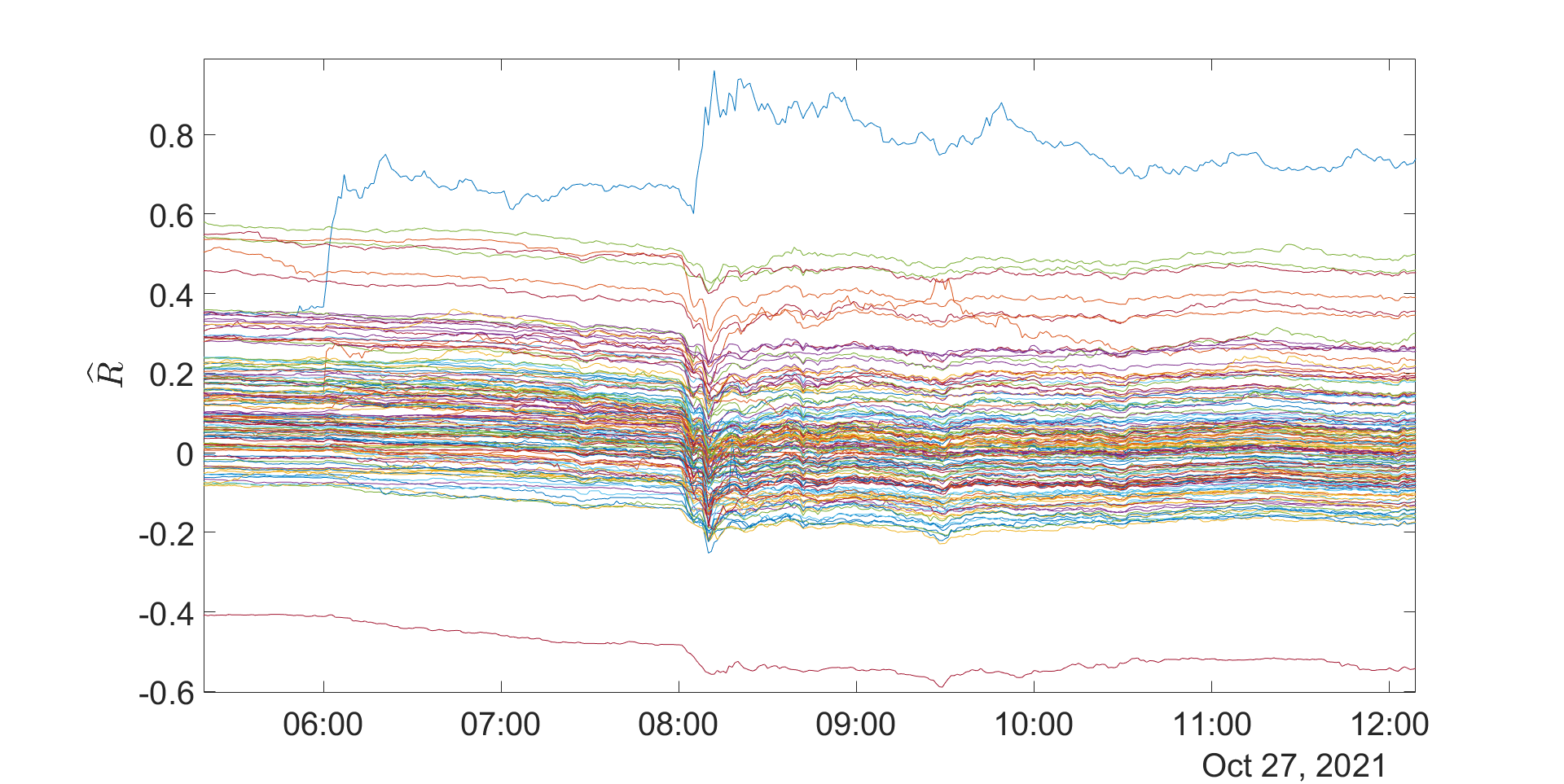}
\includegraphics[width=0.49\textwidth]{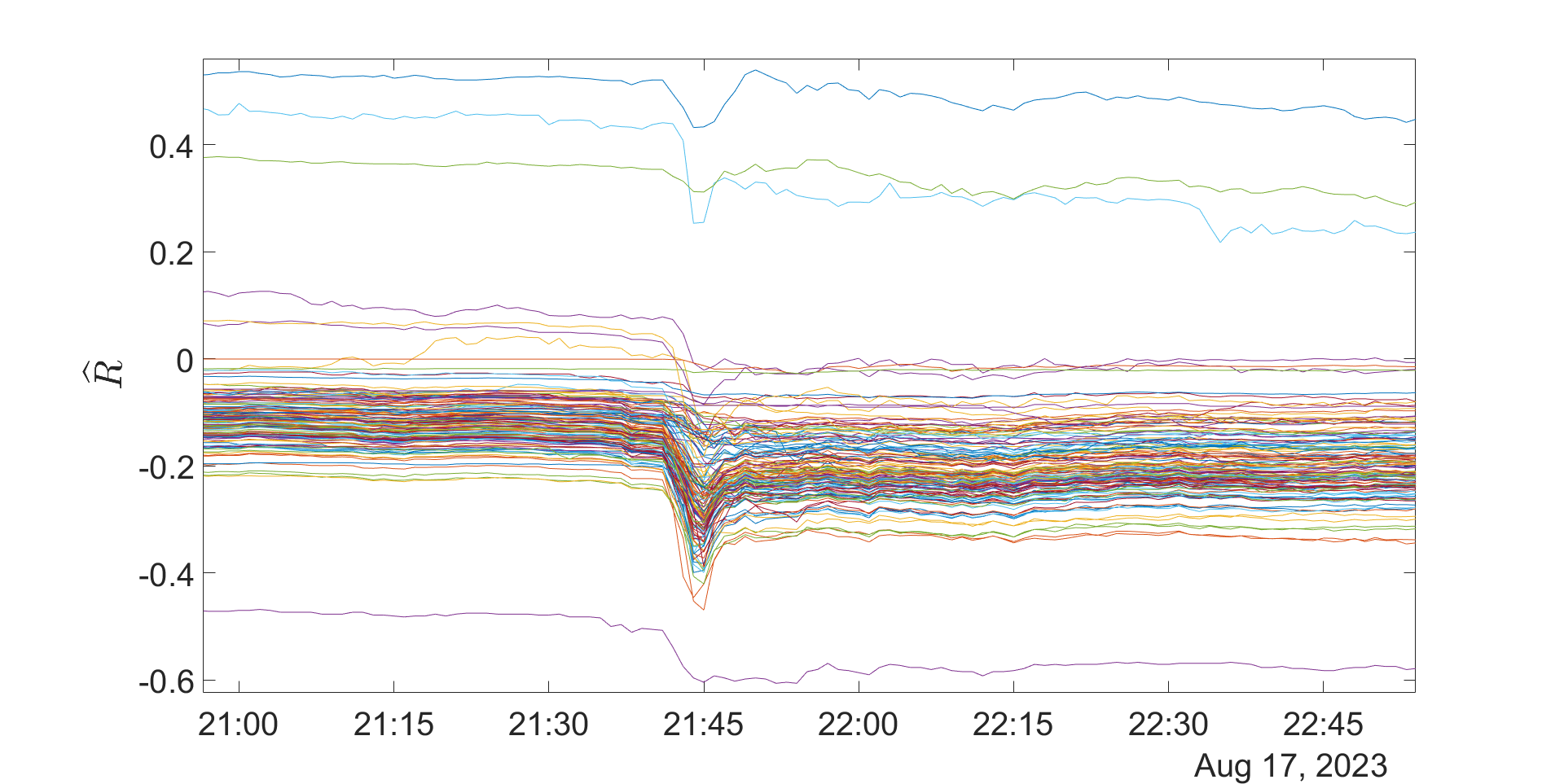}
\includegraphics[width=0.49\textwidth]{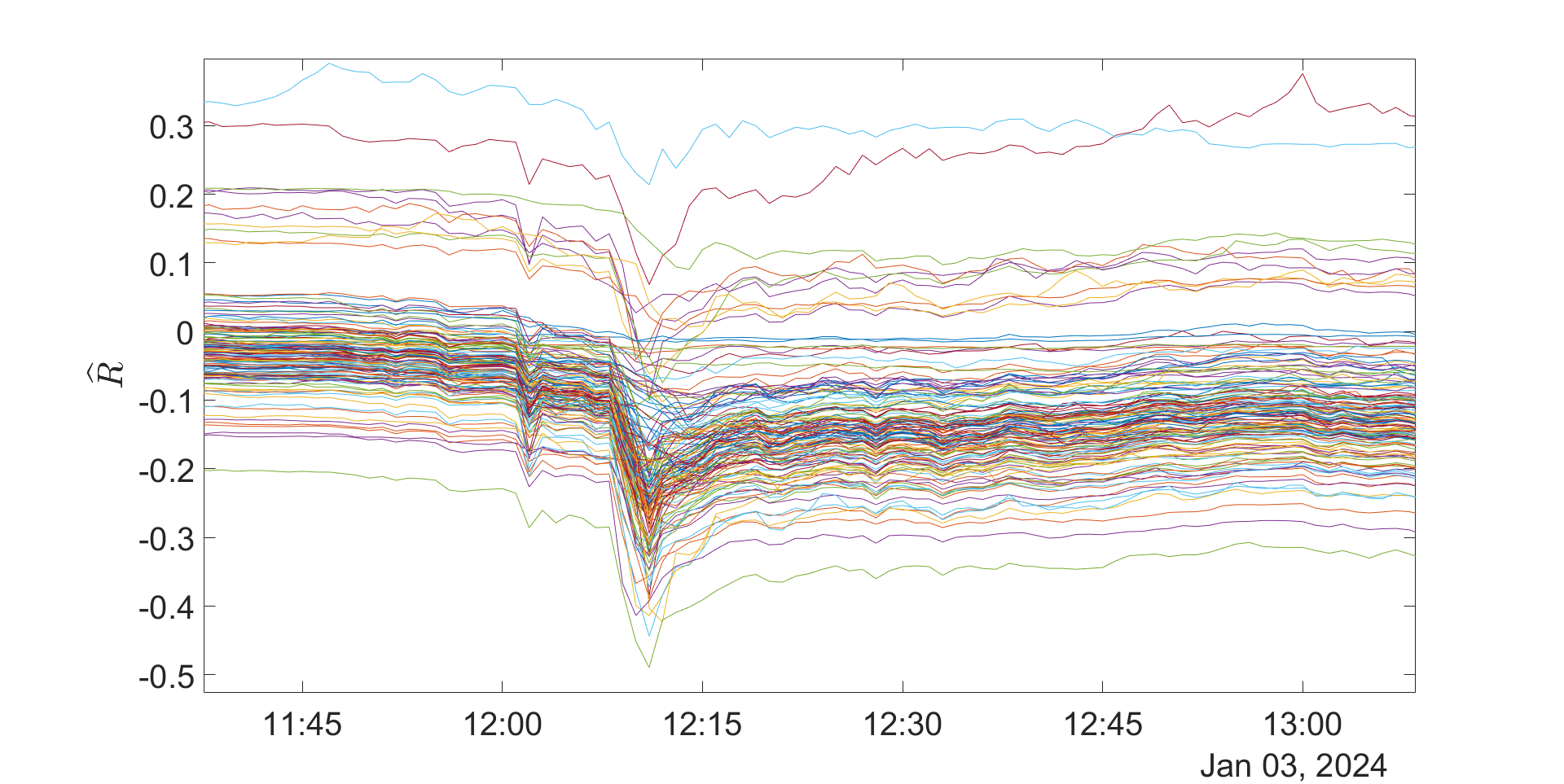}
\caption{Cumulative log-returns $\hat{R}(t)$ occurring during sample periods with large intraday drops corresponding to large difference between $q$MSTs for $q=1$ and $q=4$ presented in Figs.~\ref{fig::MST_q1_q4nr288}, \ref{fig::MST_q1_q4nr955}, and~\ref{fig::MST_q1_q4nr1094}.}
\label{fig::price_changes}
\end{figure}

To quantitatively assess when the structure of $q$MST differs the most depending on the parameter $q$, two graph distance metrics based on differences in adjacency matrices were used: DeltaCon0 denoted by $d_{\textrm{DC}0}$~\citep{KoutraD-2016a} and resistance perturbation distance denoted by $d_{\textrm{rp}1}$~\citep{MonnigND-2018a}. Their changes over time with respect to the $q$ parameter: 
\begin{eqnarray}
d_{\textrm{rp}1}[A(q=1,s=10,t),A(q=4,s=10,t)], \\ 
d_{\textrm{DC}0}[A(q=1,s=10,t),A(q=4,s=10,t)], 
\end{eqnarray}
where $A$ is the adjacency matrix for a given MST tree, are presented in Fig.~\ref{fig::rolling-windowq1_q4_s=10}c and Fig.~\ref{fig::rolling-windowq1_q4_s=10}d. It turns out that the largest values of the distance metrics can be observed in the rolling windows when the network and correlation characteristics behave as in the exceptions described above, namely when stronger correlations occur at the level of large fluctuations ($q=4$) than at the level of average fluctuations ($q=1$). These rolling windows are marked by dotted lines in Fig.~\ref{fig::rolling-windowq1_q4_s=10}. The explanation behind such situations is the price collapse of almost all cryptocurrencies within a few minutes. The trajectories of the price changes in the sample rolling windows when such a crash occurred (marked by Roman numerals in~\ref{fig::rolling-windowq1_q4_s=10}) are presented in Fig.~\ref{fig::price_changes}. There is a visible drop in all exchange rates.

% Figure 5

% Figure 6

\begin{figure}[ht!]
\centering
\includegraphics[width=0.49\textwidth]{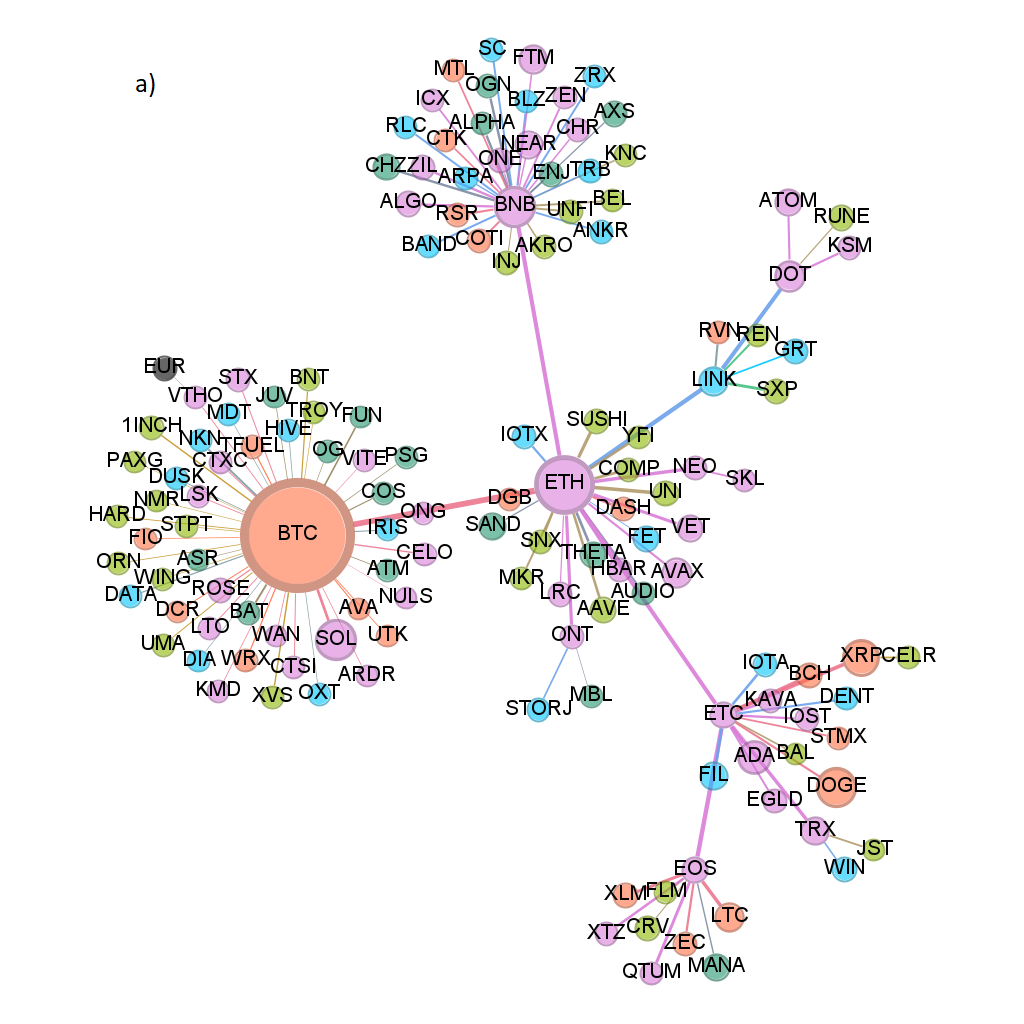}
\includegraphics[width=0.49\textwidth]{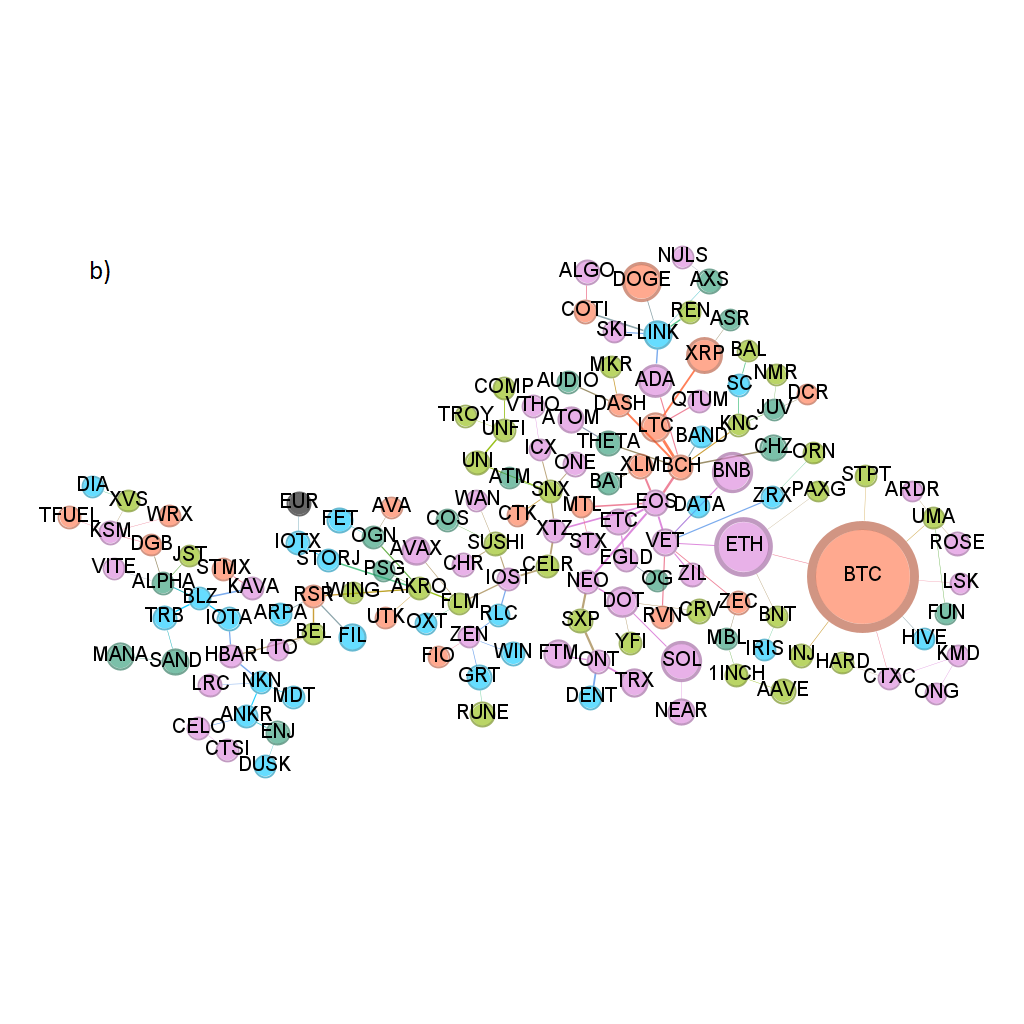}
\includegraphics[width=0.49\textwidth]{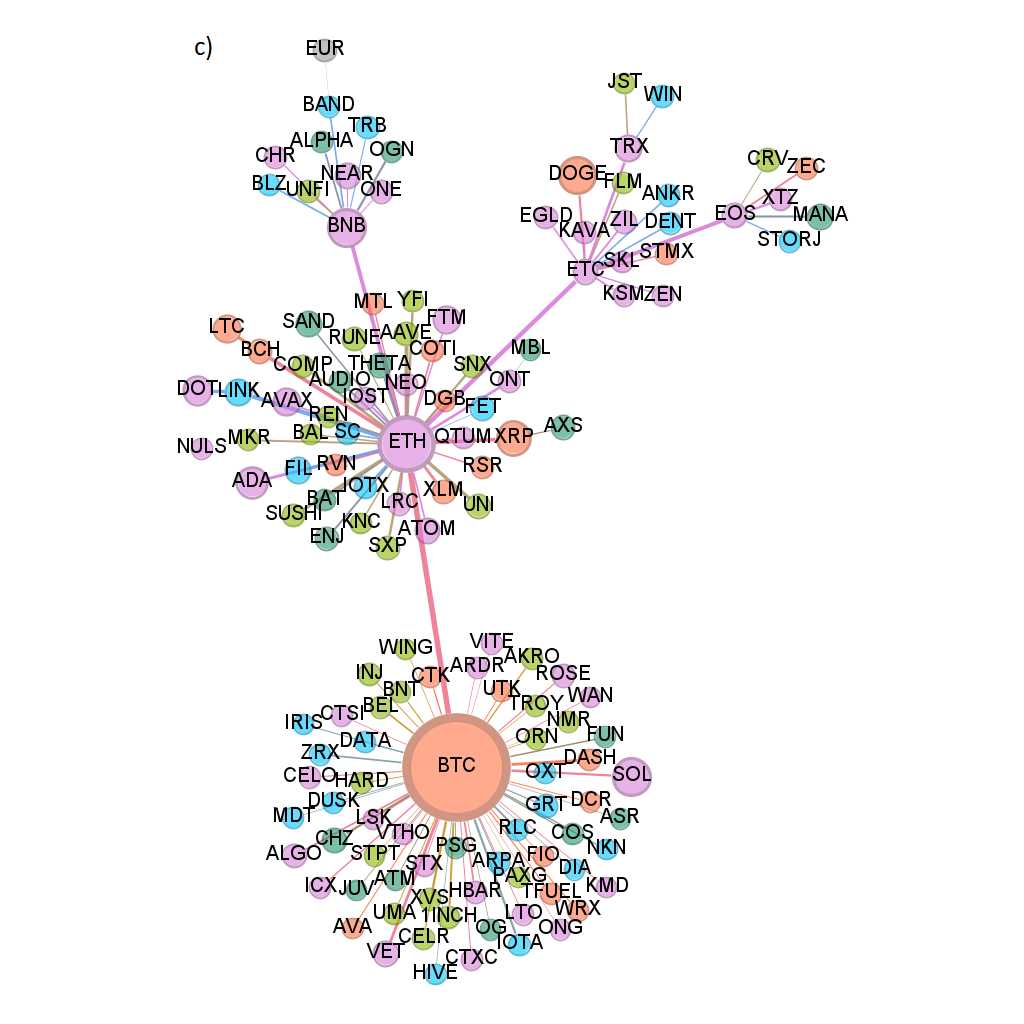}
\includegraphics[width=0.49\textwidth]{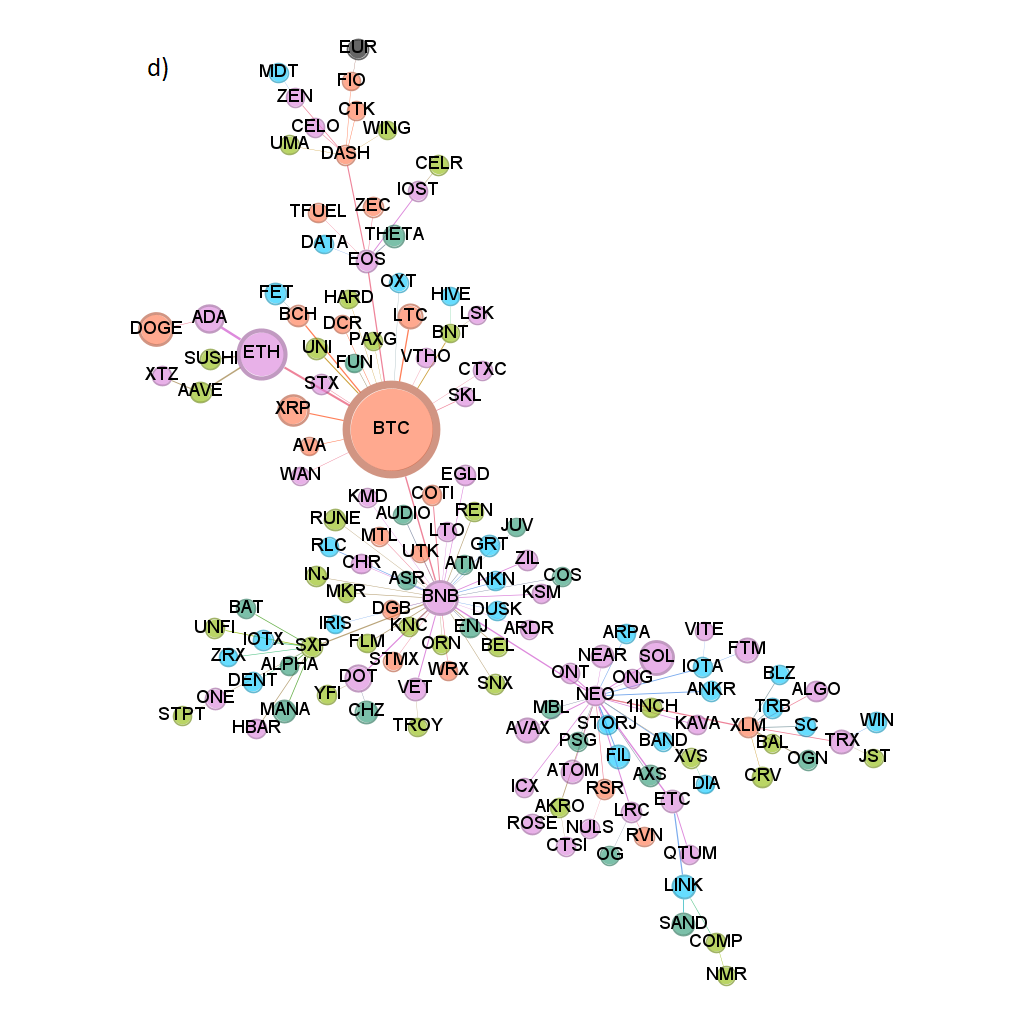}
\caption{$q$MST's calculated in a rolling window when the metrics $d_{\textrm{DC}0}$ and $d_{\textrm{rp}1}$ for $q=1$ (left - a and c) and $q=4$ (right - b and d) were among the largest: the window ending on Oct 30, 2021 (upper - a and b) and the window ending on Oct 26, 2021 (lower - c and d); these windows correspond to crashes falling out of the weekly data range.}
\label{fig::MST_q1_q4nr288}
\end{figure}

\begin{figure}[ht!]
\centering
\includegraphics[width=0.49\textwidth]{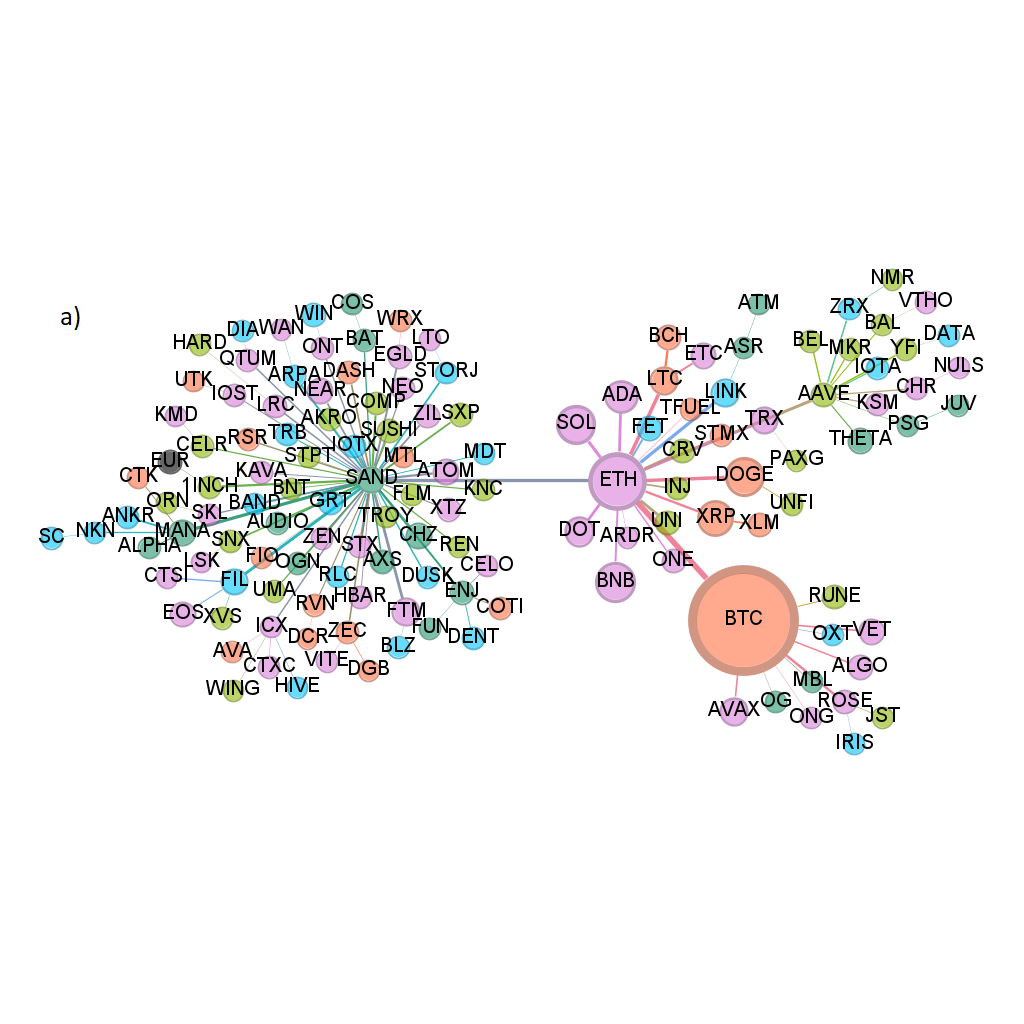}
\includegraphics[width=0.49\textwidth]{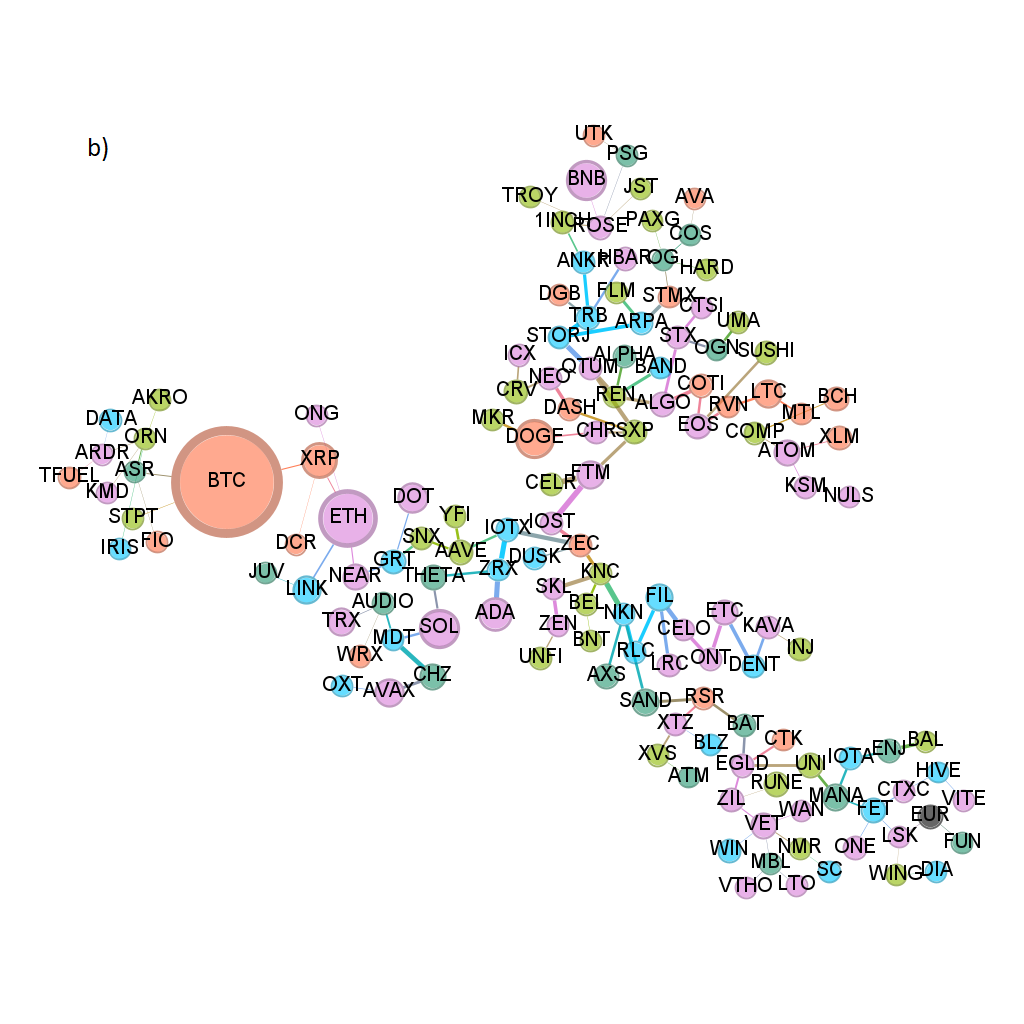}
\includegraphics[width=0.49\textwidth]{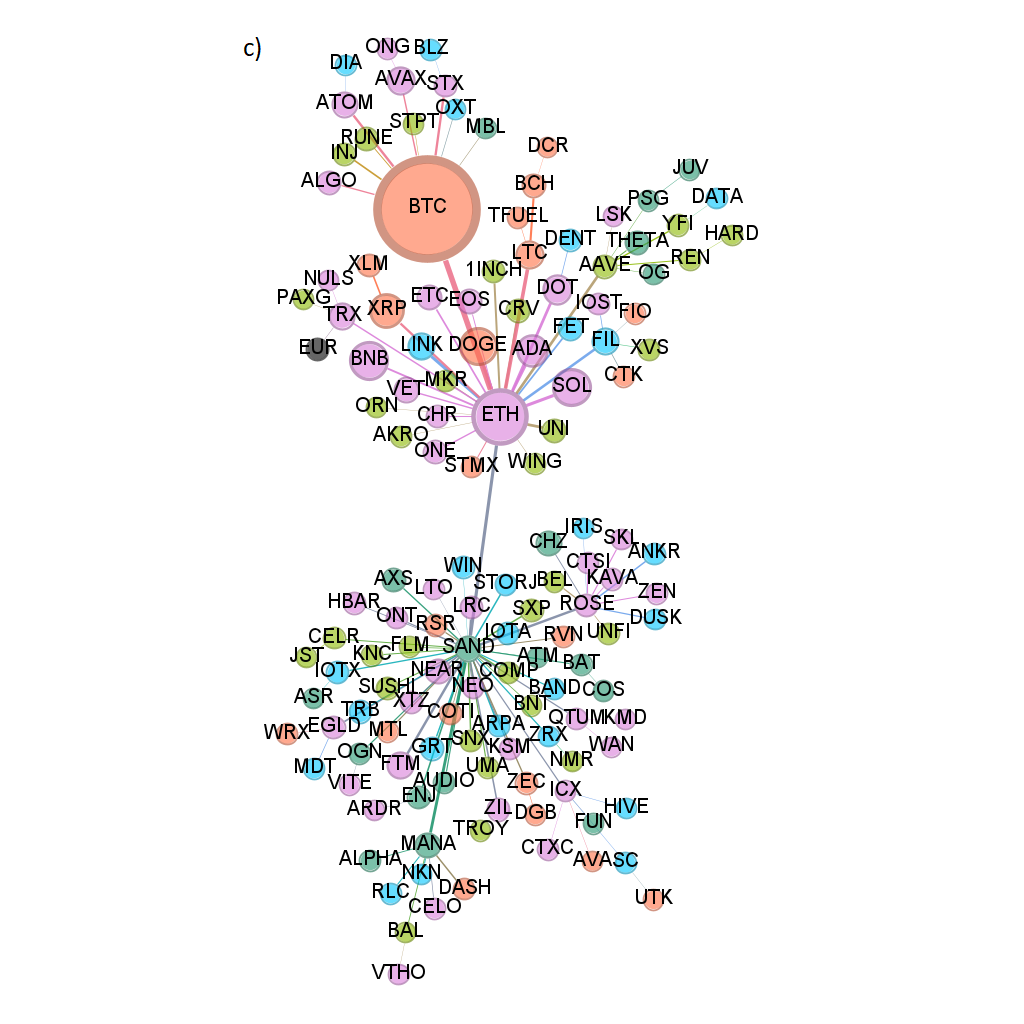}
\includegraphics[width=0.49\textwidth]{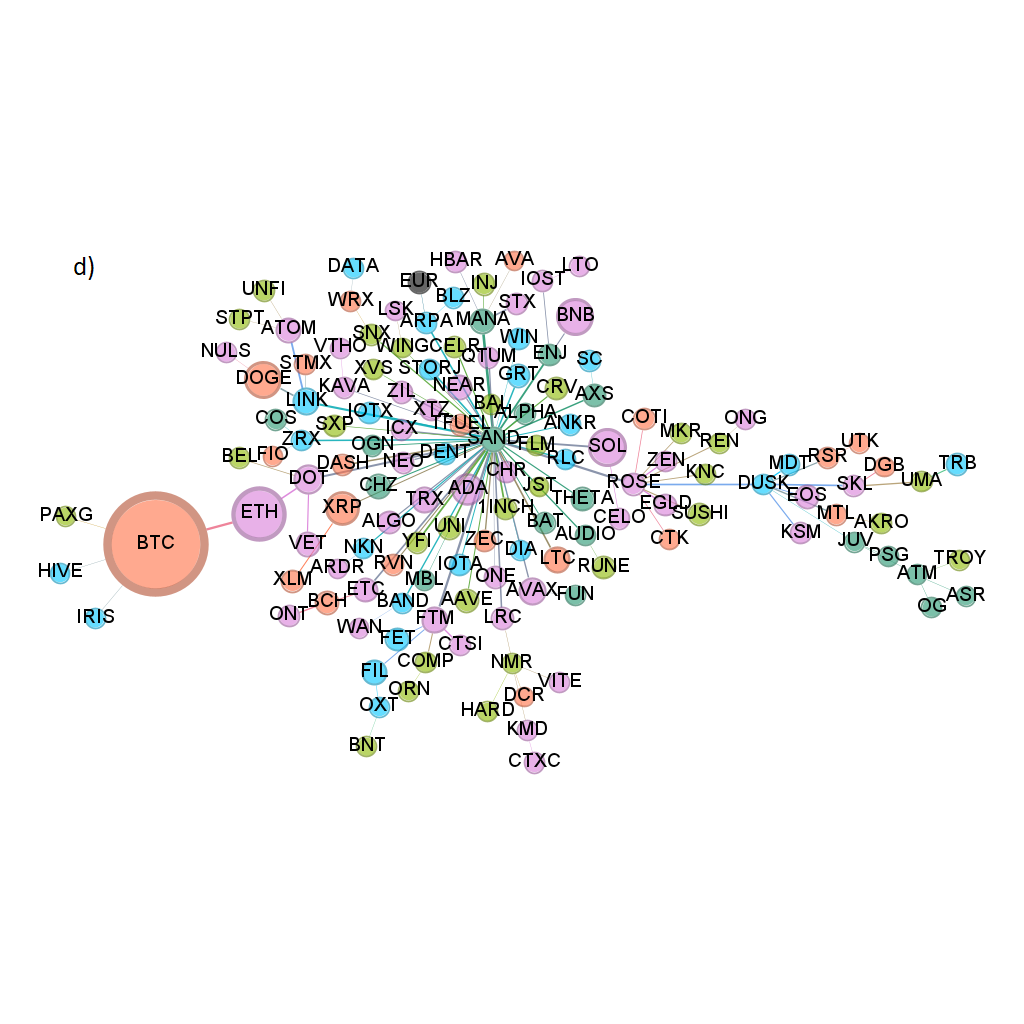}
\caption{$q$MSTs in rolling window when the metrics $d_{\textrm{DC}0}$ and $d_{\textrm{rp}1}$ for $q=1$ (left - a and c) and $q=4$ (right - b and d) were one of the largest: the window ending on Aug 24, 2023 (upper - a and b) and the window ending on Aug 5, 2023 (lower - c and d) when the crashes led to the market falling out of the weekly data range.}
\label{fig::MST_q1_q4nr955}
\end{figure}

\begin{figure}[ht!]
\centering
\includegraphics[width=0.49\textwidth]{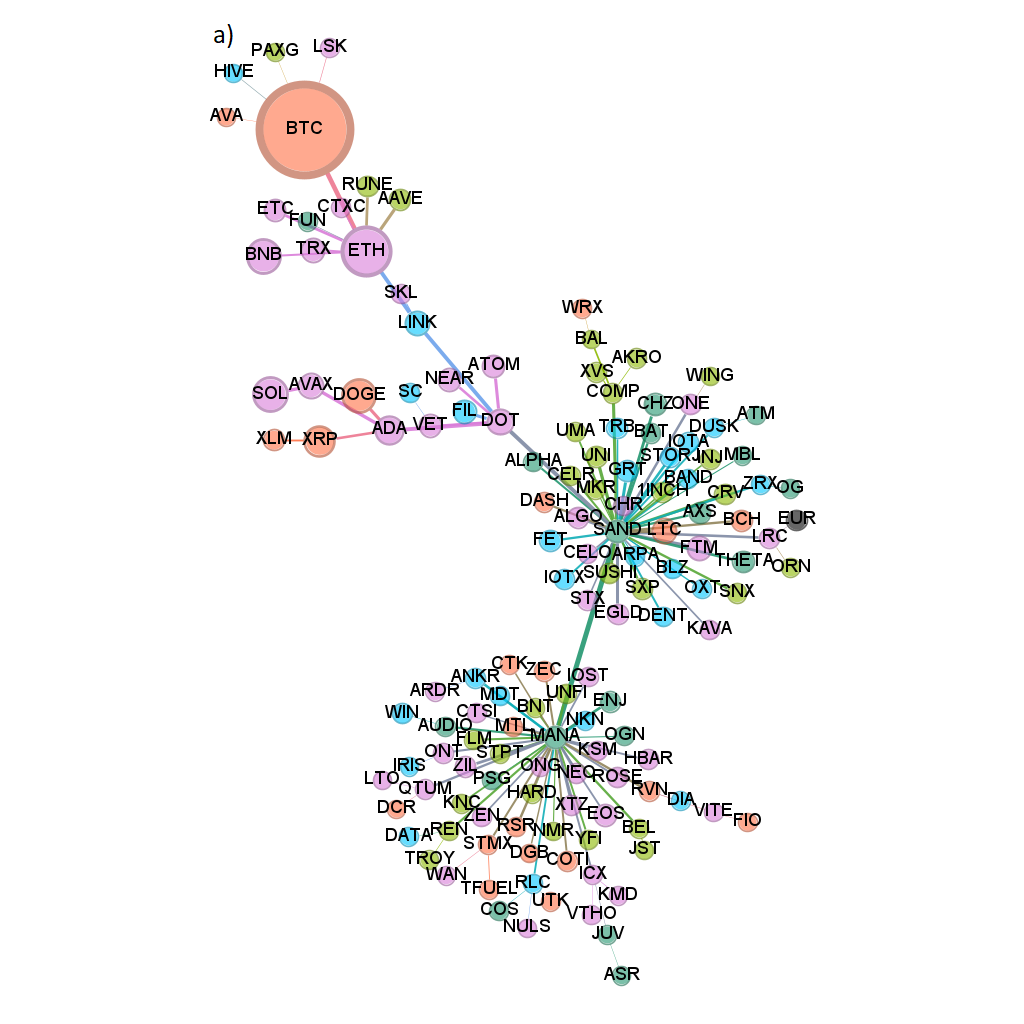}
\includegraphics[width=0.49\textwidth]{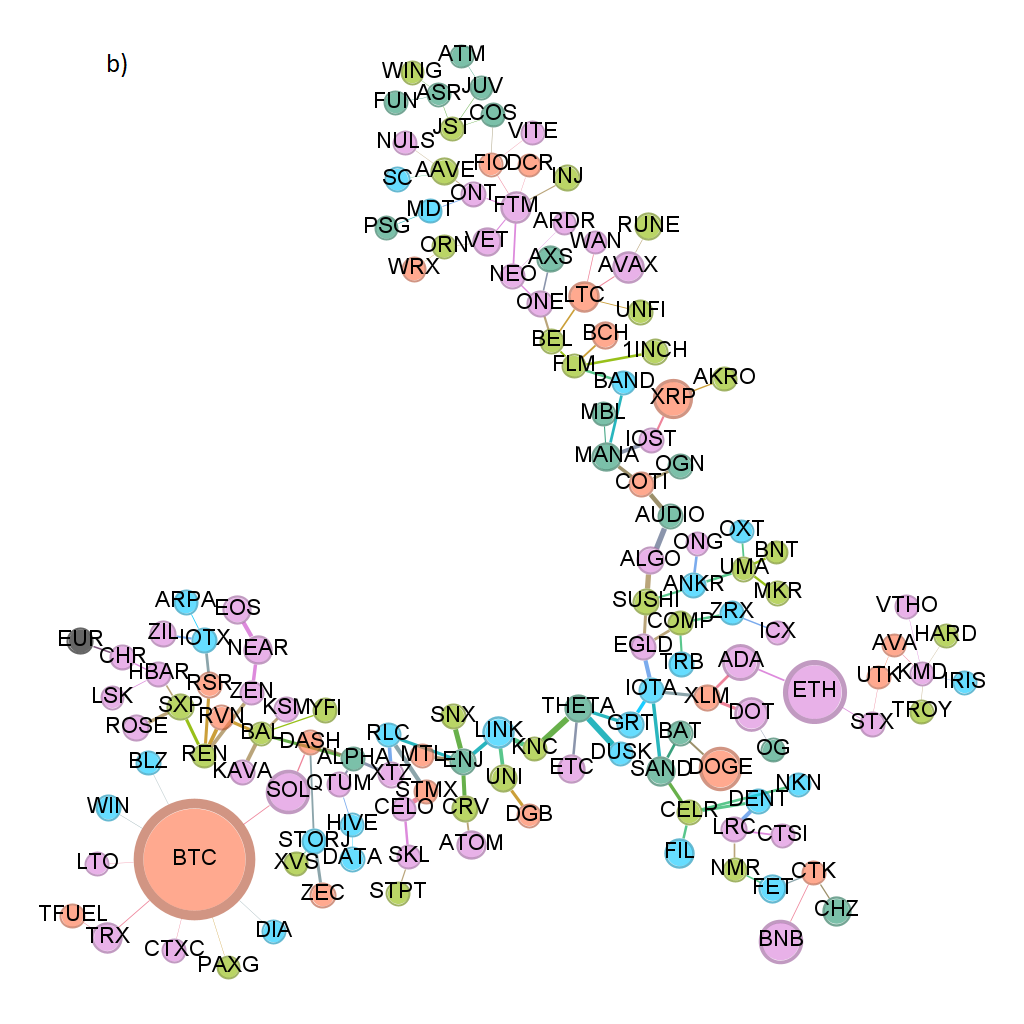}
\includegraphics[width=0.49\textwidth]{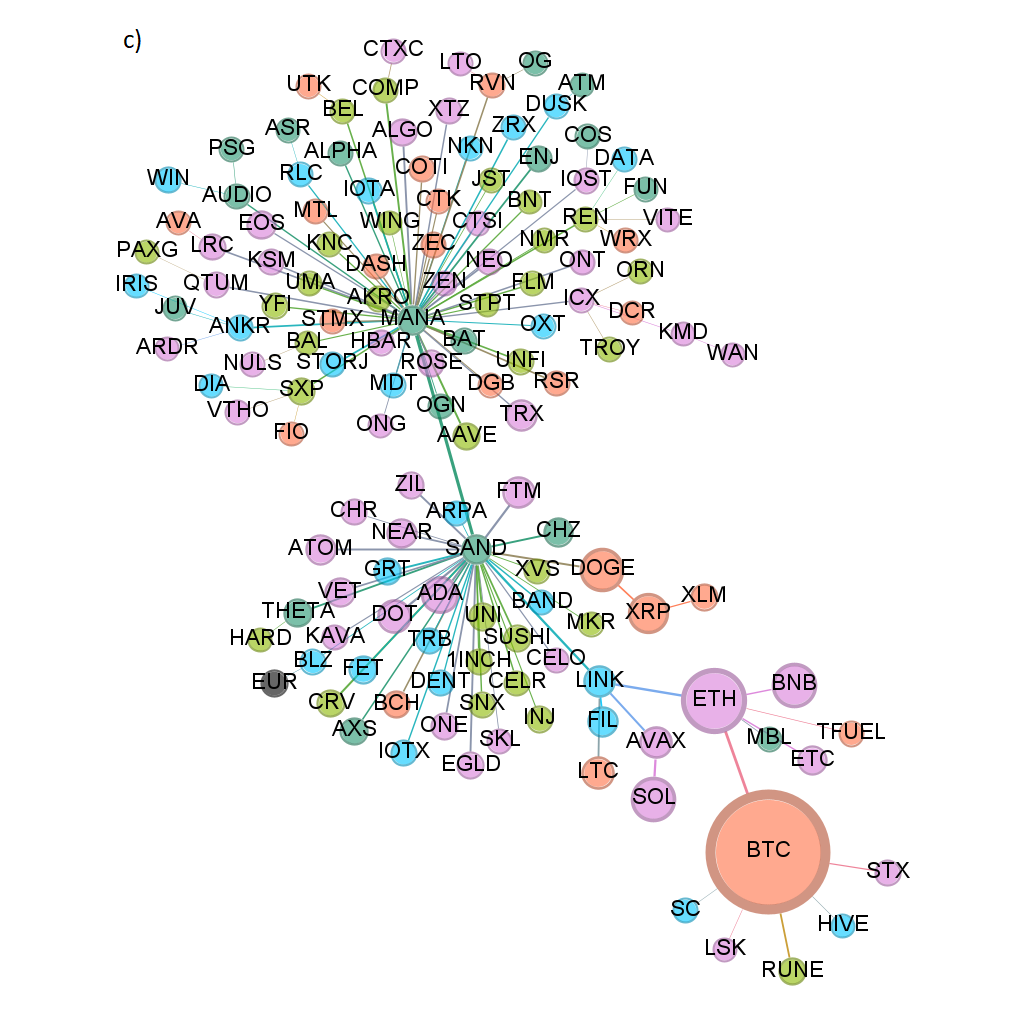}
\includegraphics[width=0.49\textwidth]{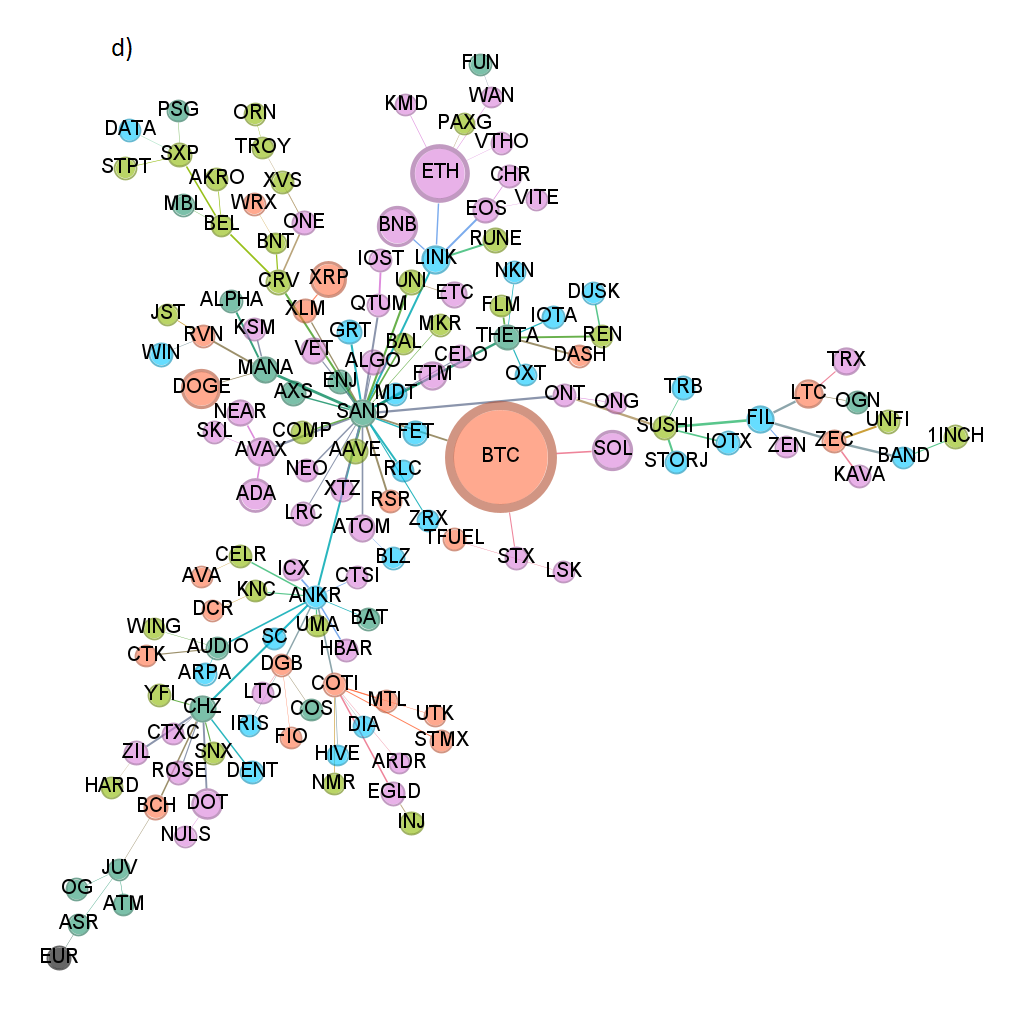}
\caption{$q$MSTs in rolling window when the metrics $d_{\textrm{DC}0}$ and $d_{\textrm{rp}1}$ for $q=1$ (left - a and c) and $q=4$ (right - b and d) were one of the largest: the window ending on Jan 10, 2024 (upper - a and b) and the window ending on Jan 11, 2024 (lower - c and d) when the crashes led to the market falling out of the weekly data range.}
\label{fig::MST_q1_q4nr1094}
\end{figure}

The $q$MSTs obtained from the date range presented in Fig.~\ref{fig::price_changes}, when the rolling windows contain crashes marked by (a) and (b) and when the crashes exceed the weekly data range marked by (c) and (d) are presented in Fig.~\ref{fig::MST_q1_q4nr288}, Fig.~\ref{fig::MST_q1_q4nr955}, and Fig.~\ref{fig::MST_q1_q4nr1094}. It is clearly visible that the structure of $q$MST for $q=4$  is entirely decentralized in the windows containing the crash, and this is not the case for $q=1$ ((a) and (b) in Figs.~\ref{fig::MST_q1_q4nr288} and~\ref{fig::MST_q1_q4nr955}). On the other hand, when there is no crash in a rolling window from which the $q$MST was obtained, the graph structure for both values of $q$ is similar ((c) and (d) in Figs.~\ref{fig::MST_q1_q4nr288} and~\ref{fig::MST_q1_q4nr955}). Such behaviour is related to the fact that, for $q=4$, the role of large fluctuations is amplified in $\rho_q(s)$, leading to stronger cross-correlations during crashes, when such large fluctuations occur. This is reflected in larger $\lambda_1$, homogeneous behaviour of the expansion coefficients, and complete decentralization of the $q$MST structure for $q=4$, because everything is strongly cross-correlated and thus behave in the same way. This effect is weaker for $q=1$, where large fluctuations are not amplified. These examples demonstrate the usefulness of the $q$MST methodology in capturing subtleties of correlation behaviour.

% Figure 7

% Figure 8

% Figure 9

Another observation is that the largest differences between graph structures for $q=1$ and $q=4$ used to occur mainly from mid-2023 to mid-2024. It indicates that the cryptocurrency market became more unstable at that time.

\subsection{Filtered correlation matrices and the corresponding $q$MSTs}
\label{sect::residual_correlations}

\begin{figure}[ht!]
\centering
\includegraphics[width=0.99\textwidth]{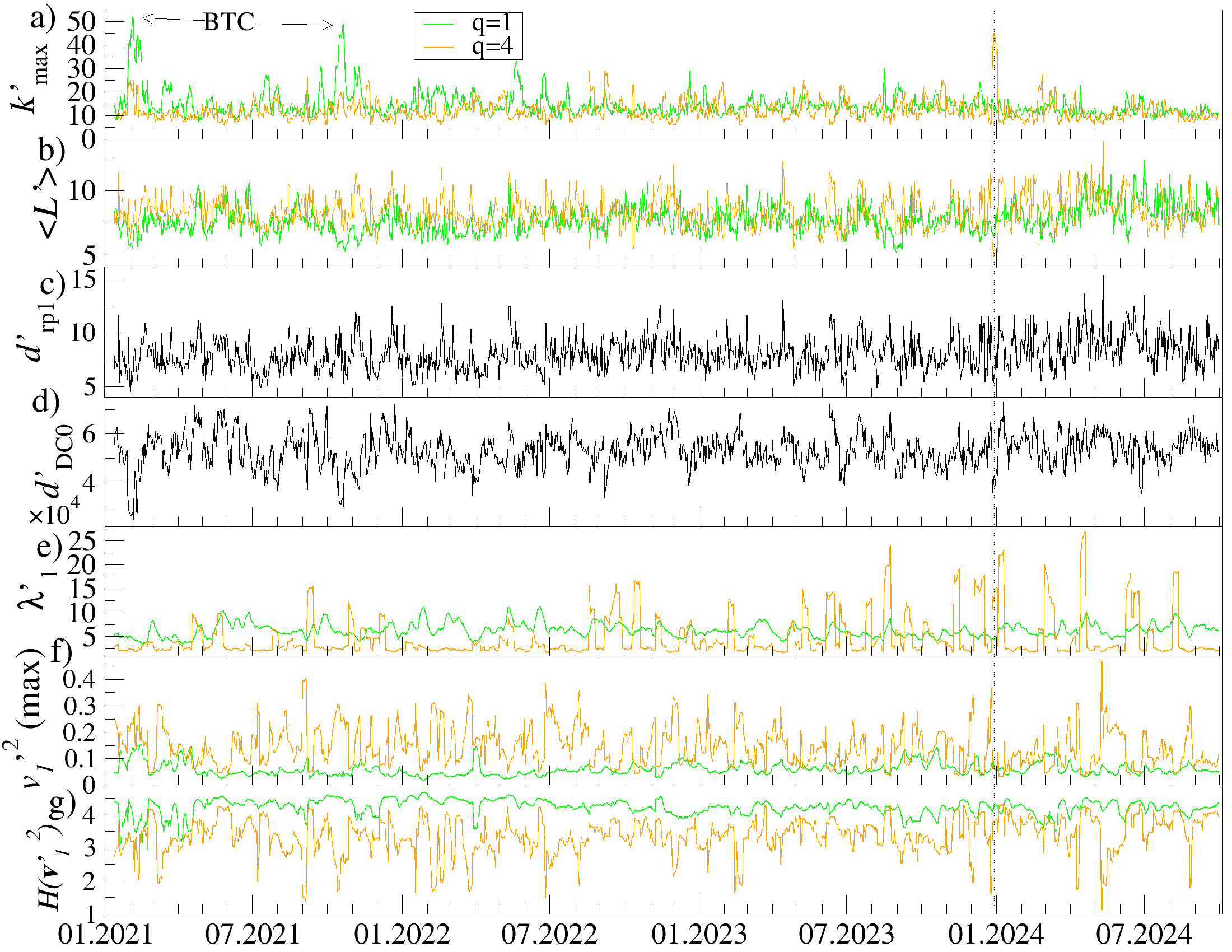}
\caption{The same measures as in Fig.\ref{fig::rolling-windowq1_q4_s=10}, but for the filtered correlation matrix ${\bf C'}$.}
\label{fig::rolling-windowfilteredq1_q4_s=10}
\end{figure}

In Sect.~\ref{sect::differences}, it was observed that during certain periods, the whole market was moving in the same direction. It was visible in an increase in the value of $\lambda_1$, which represents the so-called market factor, and the equal share of all cryptocurrencies in the eigenvector corresponding to it. This led to a reduction in the role of out-of-trend cross-correlations. To extract information about them, it is necessary to filter out the variance contribution associated with $\lambda_1$. This can be done by using a regression-based method~\citep{PlerouV-2002a,KwapienJ-2012a}:
\begin{eqnarray}
\nonumber
c_{\Delta t}^{^{\rm (i)}}(k) = a^{^{\rm (i)}} + b^{^{\rm (i)}} Z_1(k) + \epsilon^{^{\rm (i)}}(k),\\
Z_1(k) = \sum_{m=1}^N v_{1m} c_{\Delta t}^{(m)}(k),
\label{eq::regression}
\end{eqnarray}
where $Z_1(k)$ is the contribution to total variance associated with $\lambda_1$ ($k=1,...,T$) and the filtered matrix ${\bf C}'$ is constructed from the residual time series $\epsilon^{^{(i)}}(k)$ ($i=1,...,N$). It can be diagonalized by solving the problem ${\bf C}' {\bf v'}_i= \lambda'_i {\bf v'}_i$. In this subsection, the spectral properties of the residual matrix ${\bf C'}$ and the network properties of the corresponding $q$MST will be investigated in full analogy to the complete matrix ${\bf C}$ in Sect.~\ref{sect::temporal.evolution}. The first, quite obvious observation is that the cross-correlations measured with the largest eigenvalue of the filtered correlation matrix $\lambda'_1$ are weaker in all rolling windows for both values of $q$ - see Fig.~\ref{fig::rolling-windowfilteredq1_q4_s=10}e. Also, the maximum node degree presented in Fig.~\ref{fig::rolling-windowfilteredq1_q4_s=10}a is significantly smaller after correlation filtering. This results in a more decentralized structure in most windows, indicated by larger $<L>$ in Fig.~\ref{fig::rolling-windowfilteredq1_q4_s=10}b. There is also no outlier of $<L>$ observed unlike Fig.~\ref{fig::rolling-windowq1_q4_s=10}b. The smaller variation in the characteristics of the $q$MST characteristics with respect to $q$ translates into smaller values of the graph distance metrics in Fig.~\ref{fig::rolling-windowfilteredq1_q4_s=10}c and Fig.~\ref{fig::rolling-windowfilteredq1_q4_s=10}d. On the other hand, in the case of filtered correlations, there are significantly larger maximum values of the eigenvector expansion coefficient (Fig.~\ref{fig::rolling-windowfilteredq1_q4_s=10}f), which corresponds to smaller values of the expansion coefficient entropy (Fig.~\ref{fig::rolling-windowfilteredq1_q4_s=10}g) and, thus, their greater diversity, especially in large fluctuations ($q=4$). However, this takes place with much weaker correlations (smaller $\lambda'_1$). Therefore, this does not translate itself into the structure of $q$MSTs.

% Figure 10

% Figure 11

\begin{figure}[ht!]
\centering
\includegraphics[width=0.48\textwidth]{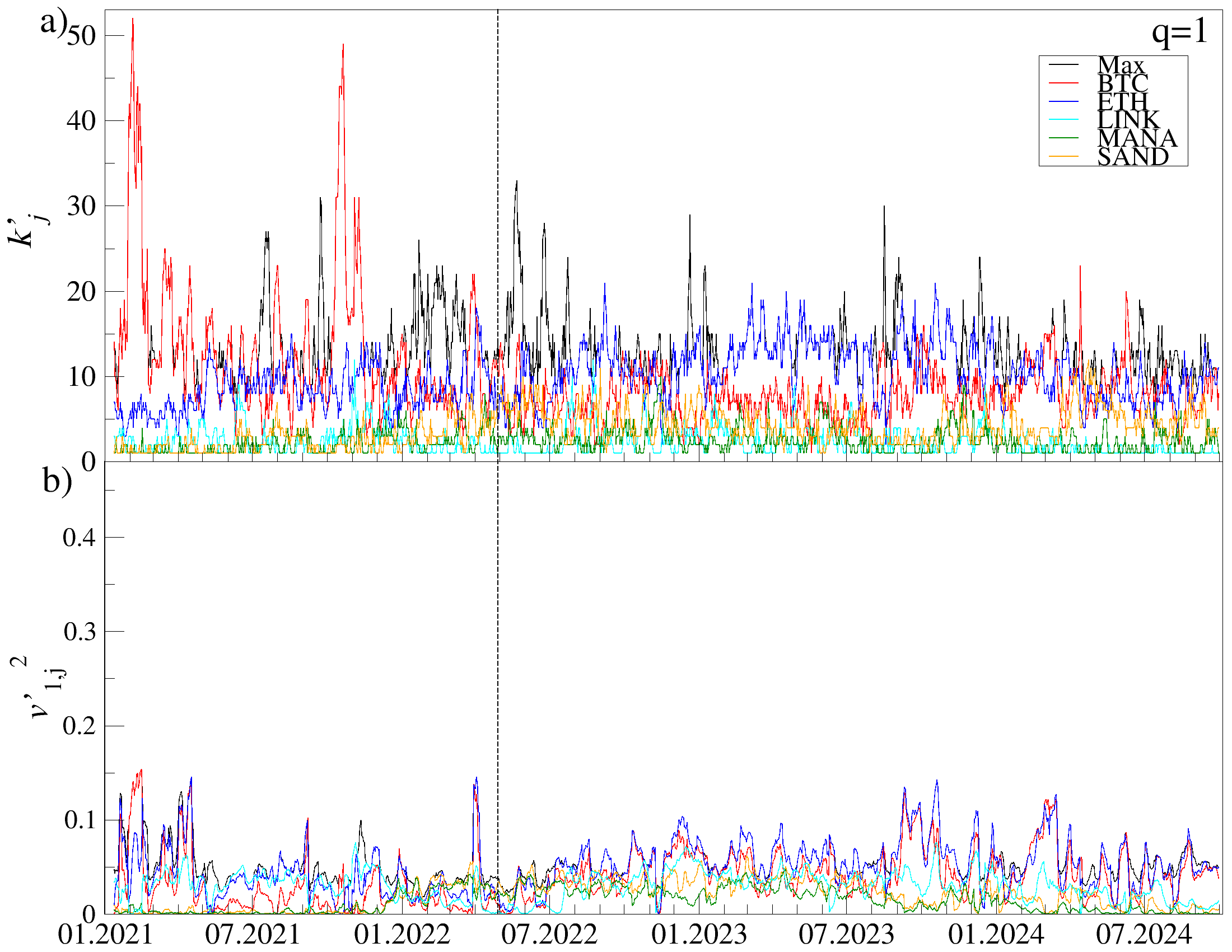}
\includegraphics[width=0.48\textwidth]{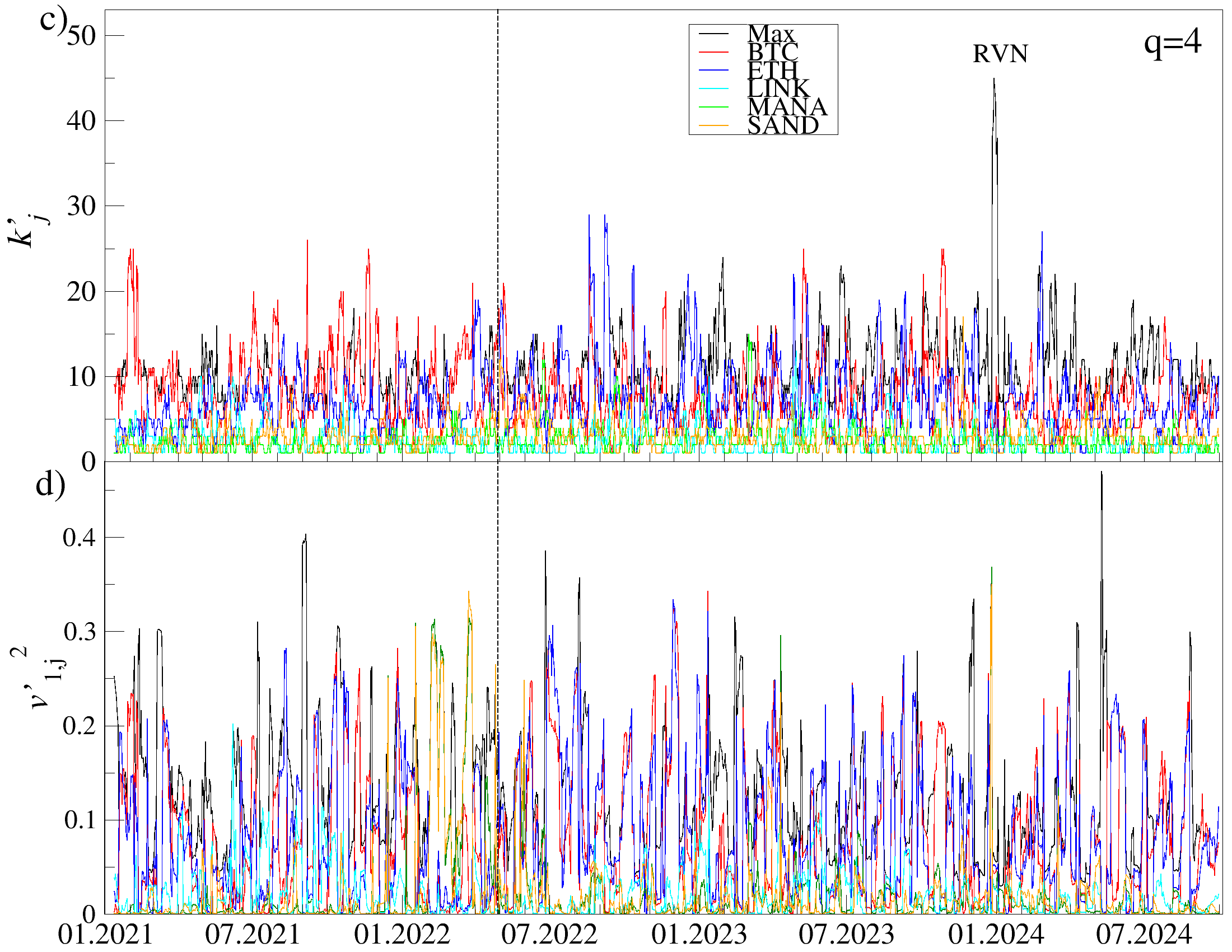}
\caption{Time evolution of the network characteristics of the $q$MSTs created from the filtered distance matrices ${\bf D'}(q=1,s=10)$ (left) and ${\bf D'}(q=4,s=10)$ (right): (a) node degree $k_j$ (cryptocurrencies that had the largest node degree in a given window were indicated) and the spectral characteristics of the $q$-dependent detrended filtered correlation matrix ${\bf C'}(q=1,s=10)$ (left) and ${\bf C'}(q=4,s=10)$ (right); (b) the squared expansion coefficients of the eigenvector ${\bf v'}^{2}_{1,j}$ associated with $\lambda'_1$ for $j$=BTC, ETH, SAND, MANA, and LINK.}
\label{fig::rolling-windowqf4_q1s=10kv}
\end{figure}

The filtering procedure has also affected the nodes with the largest multiplicity in $q$MSTs. For both $q=1$ (Fig.~\ref{fig::rolling-windowqf4_q1s=10kv}a) and $q=4$ (Fig.~\ref{fig::rolling-windowqf4_q1s=10kv}c) BTC is visible as the most connected node in 2021 until mid-2022. This effect is weaker than in the case of unfiltered correlations for $q=1$ in Fig.~\ref{fig::rolling-windowq=1s=10}a, however. On the other hand, BTC is visible as the most connected node until mid-2022 for $q=4$, which was not the case before the correlation filtering. The second difference is the absence of SAND as the node with the largest multiplicity in 2024 for both values of $q$ (Fig.~\ref{fig::rolling-windowqf4_q1s=10kv}a and Fig.~\ref{fig::rolling-windowfilteredq1_q4_s=10}b). In the case of the filtered correlations, it does not play a dominant role. It means that the appearance of SAND in 2024 as the most connected node for both values of $q$ was related to the market factor, and after filtering it out, the effect disappeared. However, the dominant role of BTC and ETH is still visible after filtering out the market factor (Fig.~\ref{fig::rolling-windowqf4_q1s=10kv}), which suggests that their role in the market correlation structure is more durable.

\subsection{Dependencies between correlation and distance matrices measures}\label{}

\begin{figure}[ht!]
\centering
\includegraphics[width=0.49\textwidth]{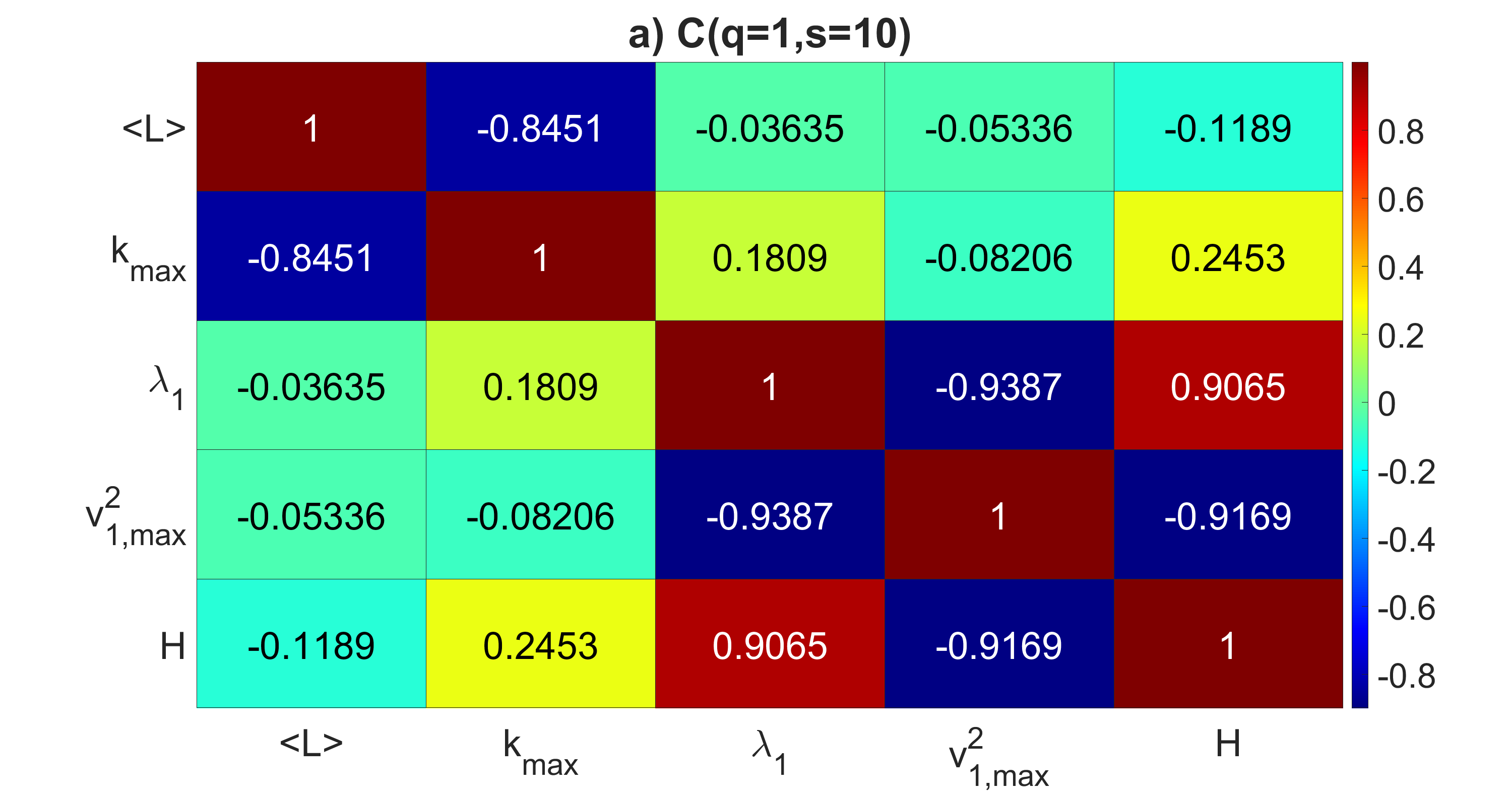}
\includegraphics[width=0.49\textwidth]{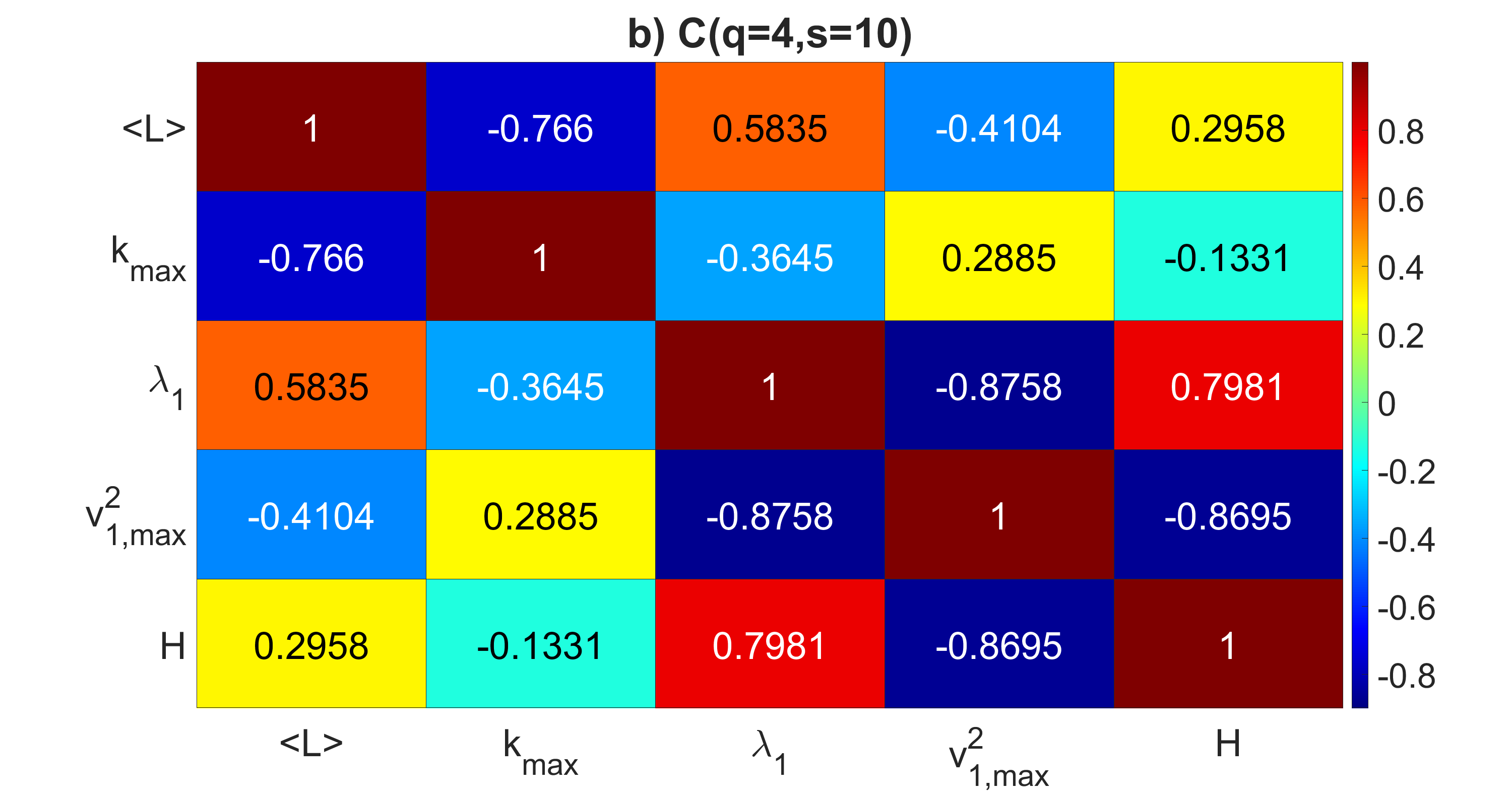}
\includegraphics[width=0.49\textwidth]{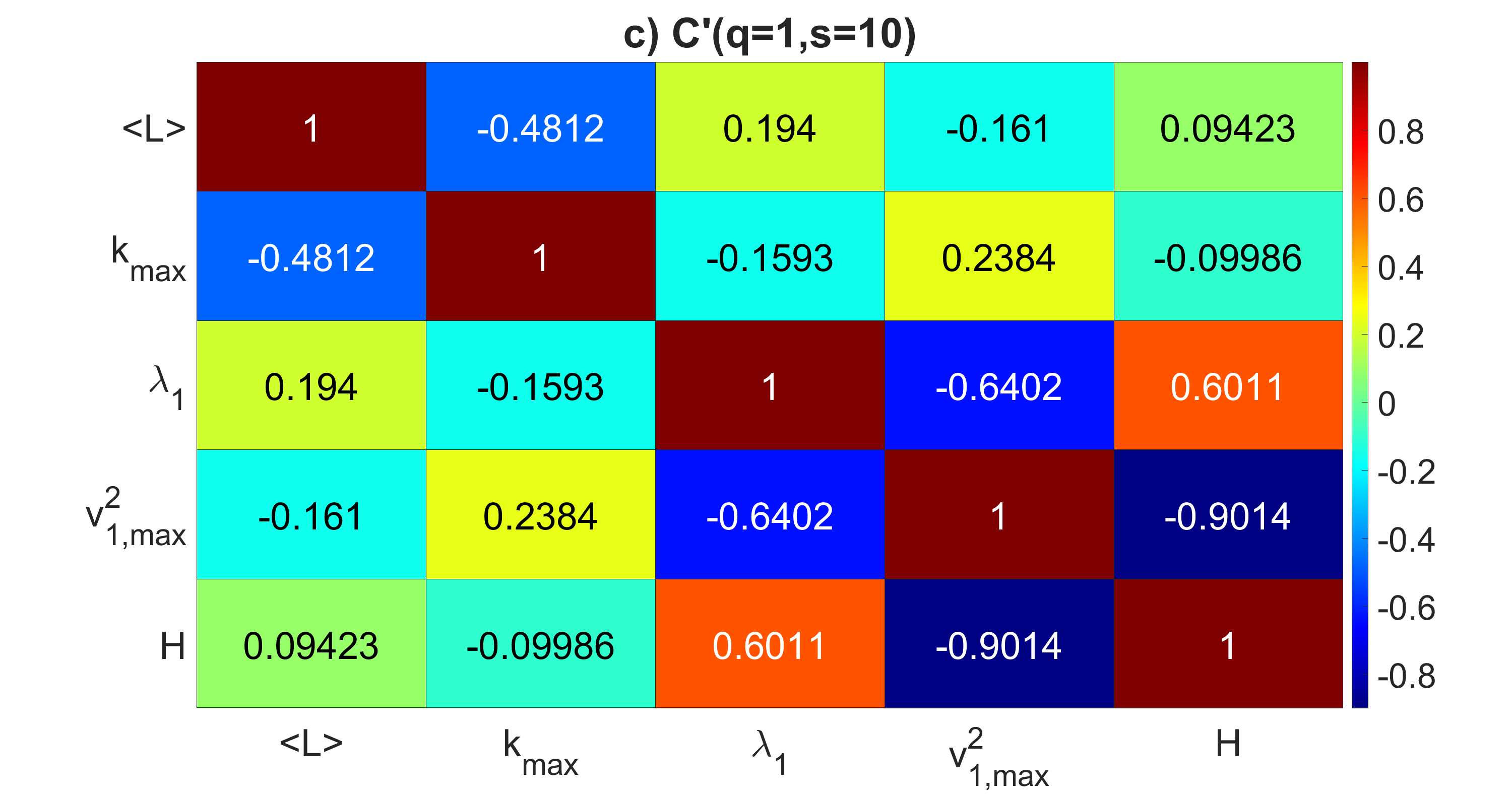}
\includegraphics[width=0.49\textwidth]{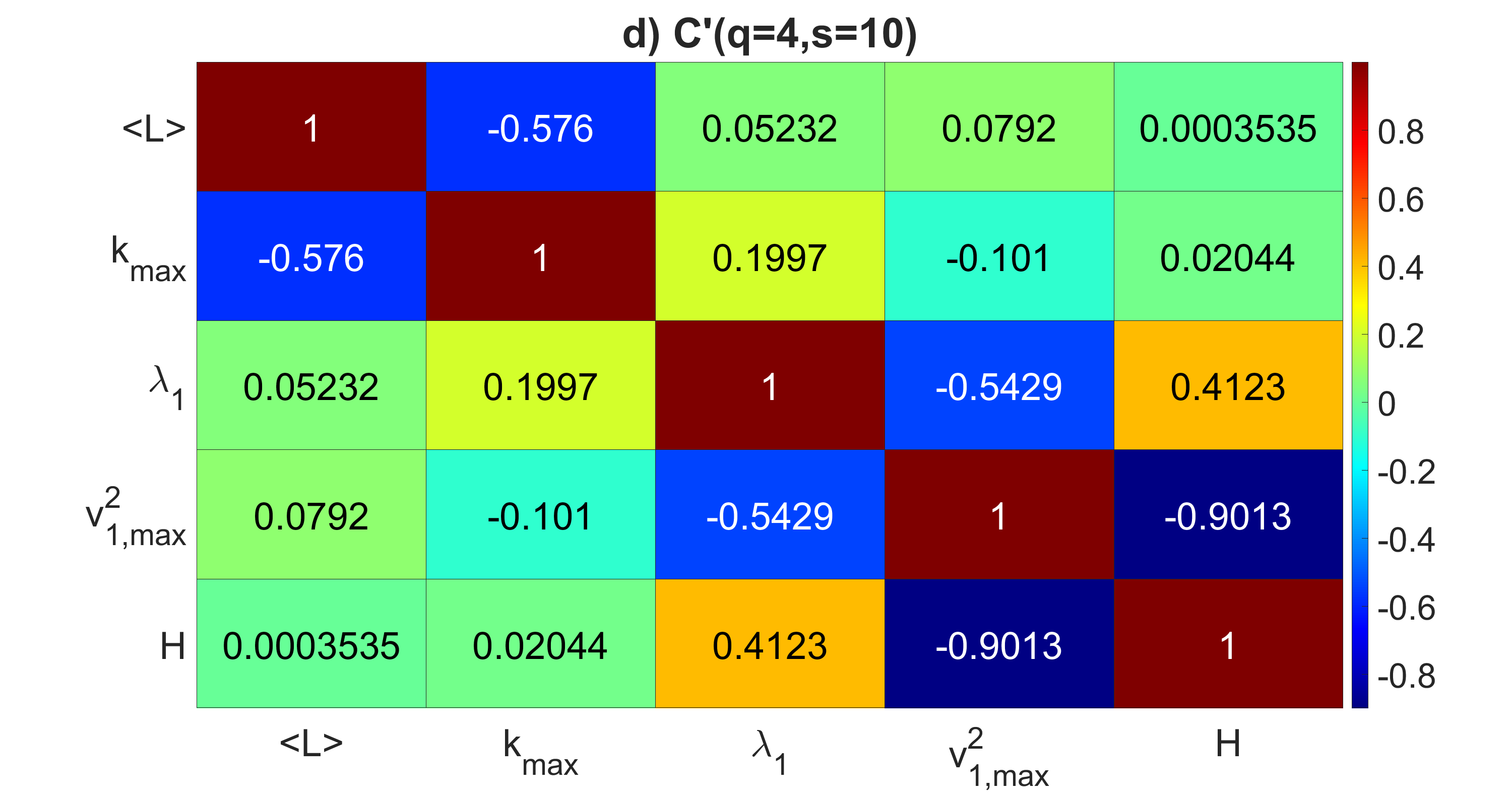}
\caption{Correlation between the spectral measures $\lambda_1$, ${\bf v}^{2}_{1,\textrm{max}}$, H(${\bf v}^{2}_{1}$) and the network measures $<L>$ and $k_{\textrm{max}}$ obtained from: (a) $C(q=1,s=10)$, (b) $C(q=4,s=10)$, (c) $C'(q=1,s=10)$, and (d) $C'(q=4,s=10)$.}
\label{fig::Cmiar}
\end{figure}

In the previous sections, the changes in the network and spectral characteristics were analyzed in rolling windows. As it can be seen from the results, the considered measures depend on each other. In order to verify quantitatively to what extent the correlations between the time series constructed from the values of the previously analyzed characteristics in each window position ($k=1,...,K$, $K=1357$ windows) were calculated using the Pearson coefficient. The results for each of the variants previously considered are presented in Fig.~\ref{fig::Cmiar}: (a) ${\bf C}(q=1,s=10)$, (b) ${\bf C}(q=4,s=10)$, (c) ${\bf C'}(q=1,s=10)$, and (d) ${\bf C'}(q=4,s=10)$.

% Figure 12

The strongest cross-correlations occur in the group of the spectral characteristics of the correlation matrix: $\lambda_1$, ${\bf v}^{2}_{1,\textrm{max}}$, H(${\bf v}^{2}_{1}$) and in the group of the network characteristics: $\langle L \rangle$ and $k_{\textrm{max}}$. They assume the highest values for medium-amplitude fluctuations ($q=1$), with their values even higher than for large fluctuations ($q=4$). The cross-correlations are correspondingly weaker after filtering out the variance associated with $\lambda_1$ (bottom panels in Fig.~\ref{fig::Cmiar}). In contrast, the cross-correlations between the metrics belonging to different groups are weak. However, in the case of large fluctuations, significant cross-correlation also occurs between the spectral characteristics of the correlation matrix and the network characteristics of $q$MSTs (Fig.~\ref{fig::rolling-windowq1_q4_s=10}b). The largest ($\approx 0.6$) is observed for $\langle L \rangle$ and $\lambda_1$. It is related to the appearance of sudden jumps in the cross-correlation level as measured by $\lambda_1$ during crashes, which influenced all the other characteristics for $q=4$ (the cases marked in Fig.~\ref{fig::rolling-windowq1_q4_s=10}).

% Figure 13

\begin{figure}[ht!]
\centering
\includegraphics[width=0.99\textwidth]{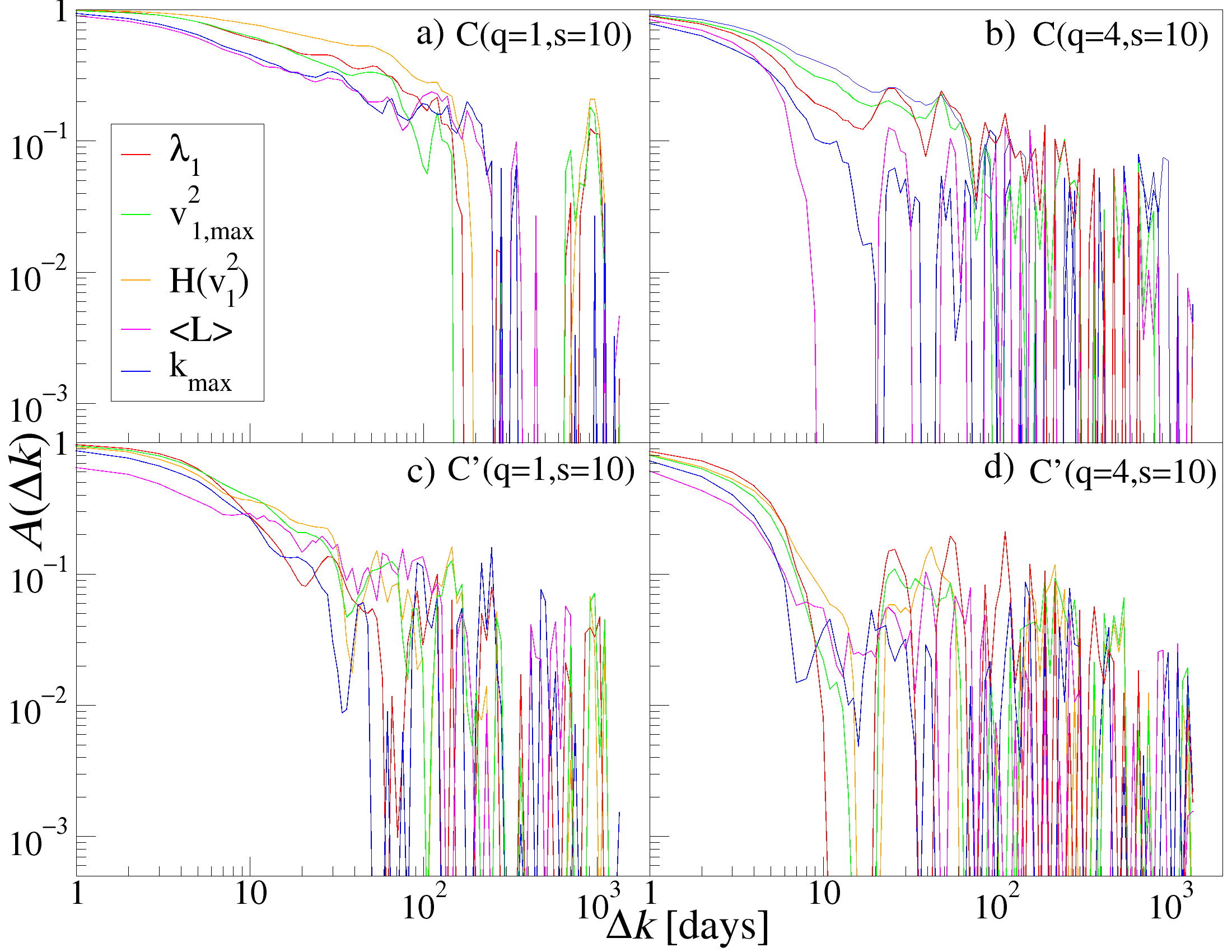}
\caption{Autocorrelation functions for $\lambda_1$, ${\bf v}^{2}_{1,\textrm{max}}$, H(${\bf v}^{2}_{1}$) $<L>$, and $k_{\textrm{max}}$ obtained from: (a) $C(q=1,s=10)$, (b) $C(q=4,s=10)$, (c) $C'(q=1,s=10)$, and (d) $C'(q=4,s=10)$.}
\label{fig::acf_miary}
\end{figure}

The cross-correlations of the analyzed characteristics seem natural due to the fact that they are based on the same correlation matrices and the $q$MSTs, but it turns out that all the considered metrics are also characterized by long-range autocorrelation, defined as
\begin{equation}
A(x,\Delta k) = { {1 \over K} \sum_{k=1}^K \left[ x(k) - \langle x(k) \rangle_i \right] \left[ x(k+\Delta k) - \langle x(k) \rangle_k \right] \over \sigma^2_x},
\label{eq::acf}
\end{equation}
where $\sigma_x$ is the estimated standard deviation of the time series $x(k)$, $\langle \cdot \rangle$ represents the estimated mean, and $\Delta k$ is the time lag in terms of rolling widows. Significant autocorrelations exceeded the obvious range of $\Delta k=6$ days, which resulted from a data overlap in 7-day windows - see Fig~\ref{fig::acf_miary}. The longest autocorrelation range occurred for the characteristics obtained for the unfiltered matrix ${\bf C}(q=1,s=10)$: $\Delta k \approx 150$ in the case of the spectral measures $\lambda_1$, ${\bf v}^{2}_{1,\textrm{max}}$, H(${\bf v}^{2}_{1}$) and $\Delta k \approx 250$ in the case of the network measures $\langle L \rangle$ and $k_{\textrm{max}}$ (Fig~\ref{fig::acf_miary}a). For $q=4$ and after filtering out the influence of the largest eigenvalue, the autocorrelation range is shorter but still significant.

Such behavior of the considered measures can be explained by a power-law decay of the volatility autocorrelation function, which is one of the stylized facts observed on all financial markets~\citep{ContR-2001a,AusloosM-2000a,KutnerR-2004a}. An interesting related fact is that the length of the power law decay in the measures is comparable with the power law decay of the ACF in the case of BTC and ETH volatility~\citep{WatorekM-2023b}. The mechanism behind it is the volatility increase because larger volatility usually results in stronger correlations, which may have an impact on the spectral and network measures, as it was shown in the previous subsections.

\section{Conclusion}
\label{concl}

This work studied the detrended cross-correlation structure of the cryptocurrency market using the $q$MST approach. In particular, it was investigated how this structure changes over time, depending on the range of fluctuation amplitude. It turned out that since May 2022, there has been a significant change in the structure of the $q$MST's. Bitcoin has ceased to be its central node, and other cryptocurrencies have taken over this role. At the same time, the $q$MST's have become more decentralized. An important part of this analysis was the identification of the dependence of the topology of $q$MST on the fluctuation amplitude. Medium-size fluctuations are more strongly cross-correlated with each other than large fluctuations. Consequently, the $q$MST graphs created from the correlation matrices with the amplified role of the largest fluctuations are more decentralized than their counterparts created from the correlation matrices with the dominant role of medium fluctuations. Moreover, the quantitative analysis of the difference between the individual cross-correlation networks depending on the fluctuation amplitude of the considered fluctuations was carried out using distance graph measures. It showed that the greatest differences occur at the time of large market events like crashes on most cryptocurrencies. During such events, the largest fluctuations were more strongly cross-correlated with each other and the $q$MST's structure were completely decentralized if these fluctuations were amplified by applying $q=4$. In the case of a focus on medium-size fluctuations with $q=2$, the graph structure was more centralized. Another general observation was that, generally, the network was becoming less centralized with increasing cross-correlation strength.

The study reported in this work illustrates the usefulness of the $q$MST concept, which allows for the selection of the fluctuation amplitude range in a network of interacting elements like assets in the financial markets. The effectiveness of this methodology was illustrated here by the example of the cryptocurrency market and the conclusions regarding its cross-correlation structure that were drawn. This may introduce novel elements for constructing optimal portfolios. The discussed spectral and network characteristics can be monitored in real time, thus the investor can use different strategies depending on the degree of correlations at the level of different fluctuations, and also detect the connections between individual cryptocurrencies and the emerging sectors in MST trees. For example, this information may support portfolio managers in applying allocation strategies: portfolios emphasizing medium-scale fluctuations would reflect stronger cross-asset correlations and thus require greater diversification to mitigate systemic risk, while portfolios constructed on large-scale fluctuations may benefit from more decentralized structures that highlight sector-specific clusters. Such insight may also be used at rebalancing time to down-weight crypto assets in the portfolio that are known to be vulnerable. In this way, $q$MST-based analysis can extend the traditional correlation-based methods by tailoring allocation, hedging, and risk management strategies to fluctuation-specific correlation patterns.  The specific use of the described dependencies in portfolio management thus emerges an inspiring issue and a promising direction for future studies. Finally, from a more general perspective, it should be noted that the same methodology can successfully be applied to many other natural and human-made complex systems.

\begin{acknowledgments}
The research was funded by National Science Centre, Poland, grant number $2023/07/X/ST6/01569$. The authors MC and MB wish to acknowledge the support from the Science Foundation Ireland under Grant Agreement No. $13/RC/2106\_P2$ at the  Research Ireland Centre at DCU. , the Research Ireland Centre for AI Driven Digital Content Technology, is funded by Research Ireland (URL: https://www.centre.ie/).
\end{acknowledgments}

\appendix

\section{List of the tickers}
\label{secA1}

\begin{longtable}{|l|l|l|l|l|l|}

\caption{\label{tab::ticker_list}Full list of the cryptocurrencies considered in this study, with the basics statistics - average volume value $\langle V_{\Delta t} \rangle$, the average number of transaction $\langle N_{\Delta t} \rangle$ and fraction of zero log-returns $\%0 R_{\Delta t}$ for $\Delta t$=1min. The classification into a given sector was made on the basis of Digital Asset Classification Standard (DACS), created by CoinDesk~\cite{coindesk}.}\\ 

\hline
\textbf{Ticker} & \textbf{Name} & \textbf{Sector} & \textbf{$\langle V_{\Delta t} \rangle$} & \textbf{$\langle N_{\Delta t} \rangle$} & \textbf{$\%0 R_{\Delta t}$} \\
\hline
\endfirsthead

\hline
\textbf{Ticker} & \textbf{Name} & \textbf{Sector} & \textbf{$\langle V_{\Delta t} \rangle$} & \textbf{$\langle N_{\Delta t} \rangle$} & \textbf{$\%0 R_{\Delta t}$} \\
\hline
\endhead

\hline
\endfoot

1INCH           & 1inch Network                 & DeFi                     & 12353       & 45          & 0.27         \\ \hline
AAVE            & Aave                          & DeFi                     & 23580       & 71          & 0.20         \\ \hline
ADA             & Cardano                       & Smart Contract Platform  & 147855      & 232         & 0.16         \\ \hline
AKRO            & Akropolis                     & DeFi                     & 3669        & 31          & 0.37         \\ \hline
ALGO            & Algorand                      & Smart Contract Platform  & 21057       & 71          & 0.20         \\ \hline
ALPHA           & Stella                        & Culture \& Entertainment & 7648        & 25          & 0.32         \\ \hline
ANKR            & Ankr network                  & Computing                & 10567       & 37          & 0.19         \\ \hline
ARDR            & Ardor                         & Smart Contract Platform  & 1475        & 9           & 0.55         \\ \hline
ARPA            & ARPA                          & Computing                & 6702        & 33          & 0.17         \\ \hline
ASR             & AS Roma Fan Token             & Culture \& Entertainment & 1730        & 11          & 0.47         \\ \hline
ATM             & Atletico Madrid Fan Token     & Culture \& Entertainment & 2589        & 12          & 0.52         \\ \hline
ATOM            & Cosmos                        & Smart Contract Platform  & 38781       & 106         & 0.09         \\ \hline
AUDIO           & Audius                        & Culture \& Entertainment & 7784        & 29          & 0.32         \\ \hline
AVA             & Travala.com                   & Currency                 & 1402        & 10          & 0.51         \\ \hline
AVAX            & Avalanche                     & Smart Contract Platform  & 78335       & 150         & 0.17         \\ \hline
AXS             & Axie Infinity                 & Culture \& Entertainment & 36413       & 72          & 0.25         \\ \hline
BAL             & Balancer                      & DeFi                     & 2380        & 11          & 0.34         \\ \hline
BAND            & Band Protocol                 & Computing                & 5845        & 24          & 0.26         \\ \hline
BAT             & Basic Attention Token         & Culture \& Entertainment & 8332        & 35          & 0.23         \\ \hline
BCH             & Bitcoin Cash                  & Currency                 & 39413       & 73          & 0.23         \\ \hline
BEL             & Bella Protocol                & DeFi                     & 4741        & 26          & 0.26         \\ \hline
BLZ             & Bluzelle                      & Computing                & 5690        & 28          & 0.30         \\ \hline
BNB             & BNB                           & Smart Contract Platform  & 271516      & 360         & 0.19         \\ \hline
BNT             & Bancor                        & DeFi                     & 2281        & 13          & 0.41         \\ \hline
BTC             & Bitcoin                       & Currency                 & 1955792     & 1693        & 0.02         \\ \hline
CELO            & Celo                          & Smart Contract Platform  & 8410        & 30          & 0.33         \\ \hline
CELR            & Celer Network                 & DeFi                     & 9515        & 33          & 0.25         \\ \hline
CHR             & Chromia                       & Smart Contract Platform  & 14296       & 45          & 0.25         \\ \hline
CHZ             & Chiliz                        & Culture \& Entertainment & 45331       & 97          & 0.23         \\ \hline
COMP            & Compound                      & DeFi                     & 8763        & 33          & 0.17         \\ \hline
COS             & Contentos                     & Culture \& Entertainment & 3960        & 25          & 0.44         \\ \hline
COTI            & COTI                          & Currency                 & 10442       & 42          & 0.25         \\ \hline
CRV             & Curve DAO Token               & DeFi                     & 22169       & 66          & 0.24         \\ \hline
CTK             & Shentu                        & Currency                 & 3219        & 22          & 0.38         \\ \hline
CTSI            & Cartesi                       & Smart Contract Platform  & 7055        & 24          & 0.30         \\ \hline
CTXC            & Cortex                        & Smart Contract Platform  & 4408        & 25          & 0.37         \\ \hline
DASH            & Dash                          & Currency                 & 10883       & 34          & 0.24         \\ \hline
DATA            & Streamr                       & Computing                & 3840        & 22          & 0.38         \\ \hline
DCR             & Decred                        & Currency                 & 970         & 8           & 0.56         \\ \hline
DENT            & Dent                          & Computing                & 13852       & 40          & 0.35         \\ \hline
DGB             & DigiByte                      & Currency                 & 3325        & 16          & 0.39         \\ \hline
DIA             & DIA                           & Computing                & 2466        & 13          & 0.46         \\ \hline
DOGE            & Dogecoin                      & Currency                 & 243222      & 345         & 0.11         \\ \hline
DOT             & Polkadot                      & Smart Contract Platform  & 105383      & 179         & 0.16         \\ \hline
DUSK            & Dusk Network                  & Computing                & 4595        & 25          & 0.33         \\ \hline
EGLD            & MultiversX                    & Smart Contract Platform  & 18389       & 53          & 0.13         \\ \hline
ENJ             & Enjin Coin                    & Culture \& Entertainment & 17924       & 51          & 0.21         \\ \hline
EOS             & EOS                           & Smart Contract Platform  & 44207       & 87          & 0.28         \\ \hline
ETC             & Ethereum Classic              & Smart Contract Platform  & 54983       & 90          & 0.21         \\ \hline
ETH             & Ethereum                      & Smart Contract Platform  & 910432      & 697         & 0.02         \\ \hline
EUR             & Euro                          & fiat currency            & 31981       & 59          & 0.49         \\ \hline
FET             & Fetch.ai                      & Computing                & 20467       & 78          & 0.19         \\ \hline
FIL             & Filecoin                      & Computing                & 62813       & 116         & 0.17         \\ \hline
FIO             & FIO Protocol                  & Currency                 & 2186        & 12          & 0.51         \\ \hline
FLM             & Flamingo Finance              & DeFi                     & 4576        & 19          & 0.38         \\ \hline
FTM             & Fantom                        & Smart Contract Platform  & 64938       & 138         & 0.10         \\ \hline
FUN             & FUNToken                      & Culture \& Entertainment & 2751        & 17          & 0.44         \\ \hline
GRT             & The Graph                     & Computing                & 20158       & 67          & 0.22         \\ \hline
HARD            & Kava Lend                     & DeFi                     & 3497        & 16          & 0.39         \\ \hline
HBAR            & Hedera                        & Smart Contract Platform  & 13672       & 43          & 0.34         \\ \hline
HIVE            & Hive                          & Computing                & 2724        & 13          & 0.36         \\ \hline
ICX             & ICON                          & Smart Contract Platform  & 5939        & 22          & 0.34         \\ \hline
INJ             & Injective                     & DeFi                     & 14880       & 51          & 0.18         \\ \hline
IOST            & IOST                          & Smart Contract Platform  & 11289       & 34          & 0.37         \\ \hline
IOTA            & IOTA                          & Computing                & 10770       & 39          & 0.22         \\ \hline
IOTX            & IoTeX                         & Computing                & 11136       & 44          & 0.20         \\ \hline
IRIS            & IRISnet                       & Computing                & 2147        & 14          & 0.41         \\ \hline
JST             & JUST                          & DeFi                     & 5775        & 15          & 0.34         \\ \hline
JUV             & Juventus                      & Culture \& Entertainment & 1871        & 11          & 0.53         \\ \hline
KAVA            & Kava                          & Smart Contract Platform  & 11703       & 38          & 0.26         \\ \hline
KMD             & Komodo                        & Smart Contract Platform  & 1774        & 14          & 0.44         \\ \hline
KNC             & Kyber Network Crystal         & DeFi                     & 5410        & 22          & 0.35         \\ \hline
KSM             & Kusama                        & Smart Contract Platform  & 10838       & 38          & 0.23         \\ \hline
LINK            & Chainlink                     & Computing                & 69537       & 138         & 0.12         \\ \hline
LRC             & Loopring                      & Smart Contract Platform  & 17362       & 51          & 0.19         \\ \hline
LSK             & Lisk                          & Smart Contract Platform  & 2792        & 13          & 0.50         \\ \hline
LTC             & Litecoin                      & Currency                 & 70616       & 142         & 0.13         \\ \hline
LTO             & LTO Network                   & Smart Contract Platform  & 2728        & 17          & 0.46         \\ \hline
MANA            & Decentraland                  & Culture \& Entertainment & 38854       & 106         & 0.11         \\ \hline
MBL             & MovieBloc                     & Culture \& Entertainment & 3761        & 19          & 0.37         \\ \hline
MDT             & Measurable Data               & Computing                & 4015        & 26          & 0.30         \\ \hline
MKR             & Maker                         & DeFi                     & 8498        & 25          & 0.33         \\ \hline
MTL             & Metal                         & Currency                 & 5867        & 24          & 0.37         \\ \hline
NEAR            & NEAR Protocol                 & Smart Contract Platform  & 45831       & 105         & 0.14         \\ \hline
NEO             & Neo                           & Smart Contract Platform  & 15854       & 40          & 0.33         \\ \hline
NKN             & NKN                           & Computing                & 5494        & 21          & 0.39         \\ \hline
NMR             & Numeraire                     & DeFi                     & 2383        & 14          & 0.46         \\ \hline
NULS            & Nuls                          & Smart Contract Platform  & 3080        & 16          & 0.42         \\ \hline
OGN             & Origin Protocol               & Culture \& Entertainment & 8763        & 33          & 0.28         \\ \hline
OG              & OG Fan Token                  & Culture \& Entertainment & 3573        & 20          & 0.37         \\ \hline
ONE             & Harmony                       & Smart Contract Platform  & 18757       & 58          & 0.21         \\ \hline
ONG             & Ontology Gas                  & Smart Contract Platform  & 2635        & 14          & 0.44         \\ \hline
ONT             & Ontology                      & Smart Contract Platform  & 12481       & 39          & 0.25         \\ \hline
ORN             & Orion Protocol                & DeFi                     & 2885        & 20          & 0.38         \\ \hline
OXT             & Orchid                        & Computing                & 3140        & 14          & 0.49         \\ \hline
PAXG            & PAX Gold                      & DeFi                     & 2727        & 6           & 0.60         \\ \hline
PSG             & Paris Saint-Germain Fan Token & Culture \& Entertainment & 3724        & 16          & 0.46         \\ \hline
QTUM            & Qtum                          & Smart Contract Platform  & 10917       & 33          & 0.24         \\ \hline
REN             & Ren                           & DeFi                     & 5274        & 23          & 0.18         \\ \hline
RLC             & iExec RLC                     & Computing                & 5950        & 24          & 0.29         \\ \hline
ROSE            & Oasis Network                 & Smart Contract Platform  & 14479       & 58          & 0.10         \\ \hline
RSR             & Reserve Rights                & Currency                 & 8500        & 38          & 0.22         \\ \hline
RUNE            & THORChain                     & DeFi                     & 30937       & 71          & 0.18         \\ \hline
RVN             & Ravencoin                     & Currency                 & 8730        & 34          & 0.21         \\ \hline
SAND            & The Sandbox                   & Culture \& Entertainment & 48390       & 108         & 0.10         \\ \hline
SC              & Siacoin                       & Computing                & 6542        & 28          & 0.38         \\ \hline
SKL             & SKALE Network                 & Smart Contract Platform  & 5869        & 27          & 0.24         \\ \hline
SNX             & Synthetix                     & DeFi                     & 10370       & 39          & 0.16         \\ \hline
SOL             & Solana                        & Smart Contract Platform  & 226651      & 363         & 0.10         \\ \hline
STMX            & StormX                        & Currency                 & 6224        & 29          & 0.27         \\ \hline
STORJ           & Storj                         & Computing                & 8109        & 34          & 0.19         \\ \hline
STPT            & STP Network                   & DeFi                     & 2724        & 17          & 0.41         \\ \hline
STX             & Stacks                        & Smart Contract Platform  & 10473       & 43          & 0.27         \\ \hline
SUSHI           & SushiSwap                     & DeFi                     & 20544       & 53          & 0.28         \\ \hline
SXP             & SXP                           & DeFi                     & 22266       & 60          & 0.26         \\ \hline
TFUEL           & Theta Fuel                    & Currency                 & 8624        & 30          & 0.38         \\ \hline
THETA           & Theta Network                 & Culture \& Entertainment & 29369       & 84          & 0.29         \\ \hline
TRB             & Tellor                        & Computing                & 10161       & 50          & 0.24         \\ \hline
TROY            & TROY                          & DeFi                     & 3198        & 19          & 0.35         \\ \hline
TRX             & TRON                          & Smart Contract Platform  & 62323       & 127         & 0.17         \\ \hline
UMA             & UMA                           & DeFi                     & 4488        & 24          & 0.31         \\ \hline
UNFI            & Unifi Protocol DAO            & DeFi                     & 8140        & 37          & 0.13         \\ \hline
UNI             & Uniswap                       & DeFi                     & 28083       & 72          & 0.19         \\ \hline
UTK             & xMoney                        & Currency                 & 2576        & 12          & 0.45         \\ \hline
VET             & Vechain                       & Smart Contract Platform  & 43606       & 99          & 0.18         \\ \hline
VITE            & Vite                          & Smart Contract Platform  & 2896        & 15          & 0.43         \\ \hline
VTHO            & VeThor                        & Smart Contract Platform  & 3757        & 24          & 0.36         \\ \hline
WAN             & Wanchain                      & Smart Contract Platform  & 1563        & 10          & 0.37         \\ \hline
WING            & Wing Finance                  & DeFi                     & 2929        & 14          & 0.49         \\ \hline
WIN             & WINkLink                      & Computing                & 21341       & 50          & 0.38         \\ \hline
WRX             & WazirX                        & Currency                 & 6646        & 26          & 0.41         \\ \hline
XLM             & Stellar                       & Currency                 & 25483       & 61          & 0.31         \\ \hline
XRP             & XRP                           & Currency                 & 228128      & 284         & 0.12         \\ \hline
XTZ             & Tezos                         & Smart Contract Platform  & 13163       & 41          & 0.35         \\ \hline
XVS             & Venus                         & DeFi                     & 8713        & 31          & 0.47         \\ \hline
YFI             & yearn.finance                 & DeFi                     & 10597       & 31          & 0.15         \\ \hline
ZEC             & Zcash                         & Currency                 & 14810       & 41          & 0.34         \\ \hline
ZEN             & Horizen                       & Smart Contract Platform  & 5913        & 25          & 0.34         \\ \hline
ZIL             & Zilliqa                       & Smart Contract Platform  & 16615       & 53          & 0.18         \\ \hline
ZRX             & 0x                            & Computing                & 5651        & 24          & 0.24         \\ \hline
\end{longtable}

\bibliography{apssamp}% Produces the bibliography via BibTeX.

\end{document}